\documentclass[prd,onecolumn,amsmath,amssymb,superscriptaddress,nofootinbib,10pt]{revtex4-2}
\usepackage[T1]{fontenc}
\usepackage{lmodern}
\usepackage{amsmath}
\usepackage{amsthm}
\usepackage{amsfonts}
\usepackage{xcolor}
\usepackage{slashed}
\usepackage{mathtools}
\usepackage{epsfig}
\usepackage[colorlinks=true,citecolor=magenta]{hyperref}

\usepackage{url}

\def\ba{\begin{eqnarray}}
\def\ea{\end{eqnarray}}

\def\be#1\ee{\begin{eqnarray}#1\end{eqnarray}}

\def\p{\partial}

\def\({\left(}
\def\){\right)}

\newcommand{\matvec}[1]{\underline{#1}}
\newcommand{\mat}[1]{\underline{\underline{#1}}}

\newcommand{\nl}{\nonumber \\}

\begin{document}

\title{Well-posedness  of minimal dRGT massive gravity}

\author{Jan Ko$\dot{\mathrm{z}}$uszek}
\email{j.kozuszek21@imperial.ac.uk}

\author{Toby Wiseman}
\email{t.wiseman@imperial.ac.uk}

\affiliation{Theoretical Physics Group, Blackett Laboratory, Imperial College, London SW7 2AZ, United Kingdom}


\begin{abstract}

Ghost-free dRGT massive gravity is a subtle theory, even at the classical level. Its viability depends on Vainshtein screening, which is an intrinsically non-linear phenomenon, and thus understanding the full non-linear dynamics of the theory is crucial. The theory was not expected to have a well-posed hyperbolic formulation as it is usually interpreted as a low energy EFT, and hence its short distance physics would be modified by higher derivative operators. Here we study a new dynamical formulation of the theory for the case of the minimal mass term. This firstly involves using a harmonic formulation of the theory, and then writing it as a first order system.
We are able to cast it in a form that is strongly hyperbolic about the Minkowski background. 
We discuss strong hyperbolicity for backgrounds close to the Minkowski solution, conjecturing that Cauchy evolution remains well-posed.
Interestingly, as part of the analysis, we find that the characteristics of the spin two graviton mode are simply governed by the inverse metric on a general background.

\end{abstract}
\maketitle
\section{Introduction}
Ghost-free dRGT massive gravity \cite{deRham:2010kj,deRham:2011qq} is the unique non-linear completion of the Pauli-Fierz mass term for linearized gravity \cite{Fierz:1939ix}. It yields arguably the most natural modification of Einstein's gravity, the addition of a mass to the graviton. Such a mass could modify the cosmological behaviour and is of great interest given the mysteries surrounding the fundamental nature of dark energy. 

Already at the classical level, the theory is subtle. Where GR propagates only two degrees of freedom, massive gravity propagates five. Moreover, the linear theory of Fierz-Pauli is discontinuous with that of GR, and produces results incompatible with observation even in the limit of small graviton mass ~\cite{vanDam:1970vg,Zakharov:1970cC}. It turns out, however, that in this limit the metric perturbation diverges, rendering a linear analysis inconsistent~\cite{Vainshtein:1972sx}. Instead, one must go to the fully non-linear theory, where a screening mechanism might arise that conceals the effects of the additional degrees of freedom on local scales (such screening is referred to as the Vainshtein mechanism, see \cite{Babichev:2013usa} for an overview). To understand the phenomenological viability of massive gravity, one must thus understand its non-linear dynamics.

Beyond the classical realm, the naive cut-off scale at which further higher dimension operators become important is very low for the theory, being constrained at less than $\sim 1000$km. However it is important to emphasize that this QFT calculation implicitly assumes perturbation theory in the metric about the Minkowski vacuum. In regions where four dimensional gravity is recovered, as discussed above, the theory will be in a non-linear regime. Instead it has been argued that in such regions one must do perturbation theory about the appropriate non-linear classical background, and a parametrically higher cut-off is obtained, which may be as high as the usual Planck scale of GR~\cite{deRham:2012ew,deRham:2013qqa}. Thus not only is understanding the classical dynamics of the theory crucial to understand consistency with classical observation, but also to understand the quantum properties of the theory.

General Relativity is famously a well-posed hyperbolic theory as shown by Choquet-Bruhat~\cite{GRWellPosed}. Well-posedness for a Cauchy problem ensures a continuous map between initial data and the subsequent evolution. 
While one might be able to find particular solutions to an ill-posed system, it cannot be regarded as a good dynamical system with an initial value problem. The importance of such well-posedness in the setting of modified theories of gravity to ensure a physically sensible Cauchy evolution problem has been emphasized in~\cite{Kovacs:2020ywu}.
We should view General Relativity as a low energy effective field theory. In general, effective field theories need not be well-posed when truncated in their derivative expansion, as we expect them to be corrected at short scales by higher derivative operators which are associated with `new' short distance physics that has been integrated out.
A good example is the UV completion given by string theory where \emph{classical} higher derivative operators arise both from dimensional reduction and also $\alpha'$ corrections.
This is the situation for relativistic viscous hydrodynamics, 
where following~\cite{Muller:1967zza,Israel:1976efz,Israel:1979wp} well-posed formulations have been given by adding new `physics' at short scales that correct the bad behaviour, as of course happens in reality when one reaches the molecular scale. If the evolution of given initial data is insensitive to changes in this short distance `completion', then this dynamics is a prediction of the effective field theory. If evolution of other initial data are sensitive to those details, this indicates a break down of classical effective field theory for that data. 
Likewise modified theories of gravity need not be well-posed, and often those involving truncated higher derivatives may not be~\cite{Papallo:2017qvl,Papallo:2017b}. Various proposals for how to complete the short distance physics to yield a well-posed Cauchy evolution problem have been given~\cite{Cayuso:2017iqc,Allwright:2018rut,Kovacs:2020ywu,Gerhardinger:2022bcw,Salo:2022, Salo:2023,deRham:2023ngf}. In the case of these higher derivative theories one may also use the fact that these higher order operators should remain small, allowing other approaches to their evolution such as~\cite{Witek:2019,Okounkova:2019,Galvez_Ghersi:2021}. Interestingly recently it has also been shown for these truncated higher derivative theories how to reformulate the theories by a change of variables to render them well-posed \cite{Figueras:2024bba}. 

The ghost-free dRGT massive gravity, even at the classical level, is generally discussed as an effective field theory, and as for relativistic hydrodynamics the folklore was that the classical theory was unlikely to be well-posed. Until recently there was no explicit dynamical formulation of the theory that allowed for numerical simulation due to the subtle constraint structure. The first such formulation was given in~\cite{deRham:2023ngf} in the case of the minimal and next-to-minimal mass terms, which are the ones believed to allow a realistic phenomenology~\cite{deRham:2014zqa,Berezhiani:2013dca,Chkareuli:2011te}. Indeed in that work numerical simulations of spherical matter collapse were performed for the minimal theory. Additional higher derivative terms were added to control poor short distance behaviour numerically, although it was unclear if this bad behaviour was due to ill-posedness or simply a numerical instability associated to the discretisation scheme used.

In this work, building on this previous dynamical formulation of~\cite{deRham:2023ngf}, we will provide a first order formulation of the classical dRGT theory for its minimal mass term. 
This theory with only minimal mass term does not exhibit a Vainshtein mechanism in spherical symmetry, and so is often taken to not be phenomenologically viable. 
Interestingly the numerical simulations of spherically symmetric collapse in~\cite{deRham:2023ngf} generically found apparently naked singularities form on the in-fall time scale. More recently in~\cite{Albertini:2024kmf} it was argued that spherically symmetric dynamics with reasonable matter are inevitably singular in the small mass limit, implying that spherically symmetric dynamics are not just in disagreement with phenomenology, but are in fact pathological.
However, as emphasized in that work, spherical symmetry is highly non-generic.
As discussed in~\cite{Renaux-Petel:2014pja}, it is it possible this minimal theory may possess a Vainshtein mechanism in generic settings with less, or no, symmetry, and hence it is still an open question whether it could agree with GR like behaviour. Thus the classical consistency of this minimal theory with observation likely requires understanding the dynamics in full generality,  beyond non-generic situations such as spherical symmetry. Our task here is precisely to initiate a study of the general dynamical structure of the dRGT theories, in order that future work may address the phenomenologically viability of this  minimal theory, and also those with more general mass terms.

By suitable modifications of a naive first order formulation which introduces auxiliary variables, we arrive at a first order system for the full non-linear theory (with no symmetry assumption) which 
we show is strongly hyperbolic when the background  spacetime is the Minkowski solution, and conjecture that dRGT with minimal mass term remains strongly hyperbolic for backgrounds in an open neighbourhood of Minkowski. We are able to prove that the local condition required for strong hyperbolicity, which also depends on the a wavevector direction, does hold for generic points and generic wavevector directions. However at a point there  may generically be a set of measure zero  of directions where our argument does not hold. This does not mean strong hyperbolicity does not hold, rather we have not been able to prove it.

While we have sufficient control over the characteristics of the theory in a neighbourhood of flat spacetime, we do not yet have a full understanding for arbitrary backgrounds, so we cannot say at what point this non-linear well-posedness breaks down, although in special cases we may analyse this. The characteristics of the theory are interesting. We will show that they comprise the two degrees of freedom of the spin-2 graviton, whose light cone is governed by the usual inverse metric. There are also three more degrees of freedom that arise from the spin-1 and spin-0 graviton modes. Interestingly these modes become birefringent on a general background, in the sense that the two degrees of freedom associated to the spin-1 graviton are no longer degenerate, and cease to share the same light-cone structure.

We begin the paper in sections~\ref{sec:dRGT} and~\ref{sec:harmdRGT} with short reviews of the ghost free dRGT gravity theory, and of its harmonic formulation for the minimal theory which was introduced in~\cite{deRham:2023ngf}. We then give a brief review of well-posedness for hyperbolic systems in section~\ref{sec:reviewwellposed}. After this in~\ref{sec:naive} we discuss a naive first order form for the harmonic minimal theory which we show is not well-posed, but introduces the variables we will use. Then we present 
a refined formulation in~\ref{sec:wellposed} and 
prove
that this is well-posed on the Minkowski background. The remaining sections~\ref{sec:zeromodes},~\ref{sec:nonzeromodes},~\ref{sec:analyticexample} and~\ref{sec:wellposedMink} then show various results about degeneracies amongst the characteristic directions of propagation on general backgrounds, and  then use these to argue that the theory
may be
 well-posed in a neighbourhood of Minkowski spacetime. 
We conclude with a brief discussion of these results and future directions.
To aid the interested reader we have produced a Mathematica notebook that performs the main computations discussed in the text~\cite{Notebook}.

\section{Brief review of dRGT gravity}
\label{sec:dRGT}

Here we will only be concerned with the classical dRGT theory.
We will follow the approach taken in~\cite{deRham:2023ngf} where rather than working with the metric as a variable, we instead use a symmetric vierbein $E_{\mu\nu}$. The starting point for the theory is then to take a reference metric, $f_{\mu\nu}$, which is naturally taken to be diffeomorphic to Minkowski spacetime.
Then from this and the vierbein the usual gravitational metric that matter couples to, $g_{\mu\nu}$, is constructed as,
\be
g_{\mu\nu} = (f^{-1})^{\alpha\beta} E_{\alpha\mu} E_{\beta\nu}
\ee
where here $(f^{-1})^{\alpha\beta}$ is the inverse of the reference Minkowski spacetime, so that $(f^{-1})^{\alpha\beta} f_{\beta\gamma} = \delta^\alpha_\gamma$. In what follows all indices will be lowered and raised using the metric $g_{\mu\nu}$. 
We will also use the notation $(E^{-1})^{\alpha\beta}$ for the inverse of the vierbein $E_{\alpha\beta}$, noting that this inverse should exist as the vierbein must be invertible to give a good metric.
Since we have used our local frame invariance to make the vierbein symmetric, we may view it as a symmetric two tensor with non-vanishing determinant (to ensure that $g_{\mu\nu}$ is a metric). We may then choose its signature to be the same as that of the metric, and thus we may regard this vierbein as itself defining a metric structure.

The action for the minimal dRGT theory may then be written as the usual Einstein-Hilbert term, together with a mass term,
\be
S = \frac{1}{16 \pi G_N} \int d^4x \sqrt{-g} \left( R[g] - m^2 \left( 2 E^\mu_{~\mu} - 6 \right) \right) + S^{(matter)}\left[ g, \Phi \right]
\ee
where $\Phi$ represents additional matter fields which (in the simplest instance) couple minimally to the metric as usual. In order to write this mass term it is crucial that there is a vierbein, and hence reference metric. Minkowski spacetime is the (maximally symmetric) vacuum solution to this theory, and $m$ then gives the physical mass of fluctuations about it. 
The most general dRGT theory can be written down with two further mass terms; the next-to-minimal mass term is quadratic in $E^{\mu}_{~\nu}$, and is explicitly given in terms of the combination $\left( (E^\mu_\nu)^2 - E^\mu_\nu E^\nu_\mu - 6\right)$ and the last term is cubic in this quantity.

We note that the minimal theory is not expected to be phenomenologically viable in spherical symmetry as it cannot provide the effective Vainshtein screening needed to recover usual GR behaviour in a small mass limit. However it may be possible to having screening in more generic settings with less, or no symmetry \cite{Renaux-Petel:2014pja}.  Including also the next-to-minimal term the theory was thought to be viable even in spherical symmetry, although recently~\cite{Albertini:2024kmf} have argued that this is not the case.
In this work we will focus only on the minimal theory for simplicity, and in future work we hope to extend discussion to include the next mass term. Since we will use the dynamical formulation in~\cite{deRham:2023ngf}, this could not be used to include the cubic mass term since it appears to have a different constraint structure \cite{Deffayet:2012nr}. In any case, that cubic term  has been argued to give rise to instabilities that put it in tension with realistic phenomenology~\cite{Berezhiani:2013dca,Chkareuli:2011te}.

While here we are providing a formal exploration of the theory, the motivation is that this can be extended to the next-to-minimal mass term precisely so that the phenomenological viability of these theories can be properly explored in the non-linear dynamical regime. Even ignoring phenomenology, these massive theories are very interesting as gravitational theories, and we may still ask very natural equations such as, "what happens when matter collapses -- do black holes form?"

The theory is diffeomorphism invariant. We may choose coordinates $x^\mu = (t, x^i)$ such that the reference metric takes the usual Minkowski form, so we have,
\be
f_{\mu\nu} = \eta_{\mu\nu} = \left( \begin{array}{cc}
-1 & 0\\
0 & \delta_{ij}
\end{array} \right) \; .
\ee
This is referred to in the literature as `unitary gauge' and we will use this throughout this work. 
If one doesn't like the notion of having a second reference metric structure on the spacetime, a more aesthetic formulation rewrites the reference metric explicitly as a diffeomorphism of Minkowski, so,
\be
f_{\mu\nu} = \frac{\partial \xi(x)^\alpha}{\partial x^\mu} \frac{\partial \xi(x)^\beta}{\partial x^\nu} \eta_{\alpha\beta}
\ee 
where now $\xi^\alpha$ are interpreted as St\"uckelberg fields, and now $\eta_{\alpha\beta}$ can be thought of as a superspace metric, rather than as a reference metric on the spacetime itself. The mass terms in the Lagrangian density can then be considered as (complicated) kinetic terms for these St\"uckelberg fields. However this formulation is, of course, classically identical to the `unitary gauge' approach we employ here.

The Einstein equation for the theory with minimal mass term is;
\be
\label{eq:dRGTEinsteinEq}
0 = \mathcal{E}_{\mu\nu} \equiv G_{\mu\nu} + m^2 M_{\mu\nu} - 8 \pi G_N T_{\mu\nu} 
\ee
where $T_{\mu\nu}$ is the conserved stress tensor of the matter fields $\Phi$, and $M_{\mu\nu}$ is given by,
\be
M_{\mu\nu} = - E_{\mu\nu} + E^\alpha_{~\alpha} g_{\mu\nu} - 3 g_{\mu\nu}
\ee
and derives from the mass term in the action, and linearizes to the Pauli-Fierz form. We note that the last part of this mass term is a cosmological constant tuned so that Minkowski spacetime  is a vacuum solution. Thus for $E_{\mu\nu} = \eta_{\mu\nu}$, and hence $g_{\mu\nu} = \eta_{\mu\nu}$ then $M_{\mu\nu} = 0$.

For pure gravity, the divergence of the Einstein equation vanishes due to the contracted Bianchi identity and conservation of the stress tensor. The same is not true of massive gravity. Instead the divergence gives the vector condition,
\be
V_\nu \equiv \nabla^\mu M_{\mu\nu} = 0 \; .
\ee
In the $m \to 0$ limit, the mass term in the Einstein equation above disappears -- however this vector condition exists for any (non-zero) mass, and it is this that is at the heart of the discontinuity seen in the linear theory~\cite{vanDam:1970vg,Zakharov:1970cC}. In the St\"uckelberg formulation we may regard this as giving equations of motion for the St\"uckelberg fields. In unitary gauge this can be thought of as a gauge condition that removes the gauge freedom of the metric (or vierbein). While the Einstein equations are second order in derivatives in the metric, or equivalently the vierbein, this vector equation is only first order and thus can be regarded as a constraint on the second order dynamics. We term this the `vector constraint'. However this is a second class constraint, meaning that it must be imposed for all times in an evolution, not simply as a constraint on initial data (as for the first class Hamiltonian and momentum constraints in GR).

In~\cite{deRham:2023ngf} a dynamical formulation of the theory was given in a 3+1 decomposition by introducing natural momenta,
\be
\label{eq:momenta}
P_i = 2 \partial_{[t } E_{i] t} \; ,\quad P_{ij} = 2 \partial_{[t } E_{i] j} 
\ee
so that morally $P_i$ and $P_{ij}$ encode the time derivatives of $E_{ti}$ and $E_{ij}$ respectively, although they are modified by spatial derivative terms. 
Since $E_{ij}$ is symmetric in~\cite{deRham:2023ngf}  only the components $P_{ij}$ with
 $i \le j$ were taken as momenta -- the other components are related by spatial derivatives, $P_{ji} = P_{ij} + 2 \partial_{[i} E_{j]t}$. We note that later in this work we will use a different convention for the components that we will choose as variables.
 It is convenient to decompose the momentum using the spatial reference metric as,
\be
P_{ij} = \frac{1}{3} \tilde{P} \delta_{ij} + \tilde{P}_{ij} \; .
\ee
If we redefine the constraint as the vanishing of the vector,
\be
\xi_\mu \equiv E_{\mu\alpha} \eta^{\alpha\beta} V_\beta
\ee
then one finds the neat expression,
\be
\label{eq:xieq}
\xi_\mu =  - 2 g^{\mu\alpha} (E^{-1})^{\beta\sigma} \partial_{[\alpha} E_{\beta] \sigma} 
\ee
and so we see that $\xi_\mu$ vanishing implies 
\be
\label{eq:vec1}
 (E^{-1})^{t i} P_i +  \frac{1}{3} (E^{-1})^{ii}  \tilde{P} = 0
\ee
from its time component, and,
\be
\label{eq:vec2}
 (E^{-1})^{t t} P_k + \frac{(E^{-1})^{t k}  }{3} \tilde{P}  = (E^{-1})^{t i} \tilde{P}_{ki}  + 2 (E^{-1})^{i t} \partial_{[i} E_{k] t} + 2 (E^{-1})^{i j} \partial_{[i} E_{k] j}
\ee
from its spatial components, so that suitably near flat spacetime, this linear constraint on the momenta may be solved for $\tilde{P}$ and $P_i$.

We now note that no momentum variable has been assigned for $E_{tt}$. This variable is determined algebraically by an additional constraint equation, the `scalar constraint', which results when the trace of the Einstein equation is combined with the gradient of the vector constraint as,
\be
\label{eq:scalarconstraint}
0 = \mathcal{S} \equiv - R + 2 \nabla^\mu \xi_\mu + m^2 M^\mu_{~\mu} - 8 \pi G_N T \; .
\ee
This can be written in terms of the vierbein as,
\be 
\label{eq:scalarconstraintexpr}
A^{\alpha\beta\gamma\mu\nu\rho} \partial_{[\alpha} E_{\beta]\gamma} \partial_{[\mu} E_{\nu]\rho} + 3 m^2 \left( E^\mu_{~\mu} - 4 \right) = 8 \pi G_N T
\ee
where the tensor $A^{\alpha\beta\gamma\mu\nu\rho}$ is given by,
\be
\label{eq:Atensor}
A^{\alpha\beta\gamma\mu\nu\rho} = \eta^{\gamma\rho} g^{\alpha [\mu} g^{\nu] \beta} - 2 (E^{-1})^{\rho [\alpha} g^{\beta][\mu} (E^{-1})^{\nu ] \gamma} 
+ 4 (E^{-1})^{\gamma [\alpha} g^{\beta][\mu}   (E^{-1})^{\nu ]\rho} \; .
\ee
Due to the antisymmetry in the derivative terms, we see that there are no time derivatives of $E_{tt}$ (assuming a minimal coupling of matter to the metric). However there are spatial derivative terms, but once one replaces time derivatives with the momenta~\eqref{eq:momenta} these are all subsumed into the momentum terms since $\partial_i E_{tt} =  \partial_t E_{ti}-  P_i $. Thus finally writing this in terms of the vierbein components and momenta, $E_{tt}$ enters only algebraically.

Decomposing $E_{ij} = \frac{1}{3} \tilde{E} \delta_{ij} + \tilde{E}_{ij}$, we may then regard the second order dynamical degrees of freedom as those with unconstrained momenta, so the five components $\tilde{E}_{ij}$ with their momenta $\tilde{P}_{ij}$, which encode the five massive graviton degrees of freedom. 
The evolution equations for $\tilde{P}_{ij}$ can be taken from the spatial components of the Einstein equations, $\mathcal{E}^i_{~j}$, minus one linear combination (naturally the trace part).
While the vector and scalar constraints determine $\tilde{P}$, $P_i$ and $E_{tt}$ algebraically, these evolution equations from $\mathcal{E}^i_{~j}$ do contain time derivatives of these variables. Hence these must be eliminated by computing the time derivative of the vector and scalar constraints which then determine these quantities $\dot{\tilde{P}}$, $\dot{P}_i$ and $\dot{E}_{tt}$.
Once these are eliminated then  the traceless part of $\mathcal{E}^i_{~j}$ determines $\dot{\tilde{P}}_{ij}$. 
In addition there is an auxiliary first order system formed from the definitions of the momenta in~\eqref{eq:momenta} for $P_i$ and $\tilde{P}$, together with the vector constraint determining these~\eqref{eq:vec1} and~\eqref{eq:vec2}, which then gives  evolution equations for $E_{ti}$ and $\tilde{E}$.

Recall that of the Einstein equations, we directly solve the scalar constraint and the spatial components minus one linear combination, so in total 6 linear combinations of components -- thus 4 linear combinations are not directly solved for, but instead the vector constraint is solved which involves the divergence of the Einstein equations, and hence their derivatives. How do we ensure that these remaining 4 linear combinations of components actually vanish (rather than just derivatives of them)? As discussed in~\cite{deRham:2023ngf}, just as for GR, the vanishing of these implies the initial data must satisfy the Hamiltonian and momentum constraints which, as usual, are given by the Einstein equation components $\mathcal{E}^t_{~\mu}$ (and can be taken to be the 4 linear combinations we do not directly impose).

\section{Harmonic formulation of dRGT gravity}
\label{sec:harmdRGT}

Here we will employ a harmonic formulation of the minimal dRGT massive gravity, also given in~\cite{deRham:2023ngf}. Let us begin by reviewing a similar formulation for GR.

\subsection{A harmonic formulation for GR}

With pure gravity, a harmonic formulation may be given by introducing a reference metric $\bar{g}_{\mu\nu}$ with connection $\Gamma^\alpha_{~\mu\nu}$, and constructing a vector field (see for example~\cite{Wiseman:2011by}),
\be
v^\alpha = g^{\mu\nu} \left( \Gamma^\alpha_{~\mu\nu} - \bar{\Gamma}^\alpha_{~\mu\nu} \right)
\ee
and then taking the `harmonic Einstein equation',
\be
R_{\mu\nu}  - \nabla_{(\mu} v_{\nu)}  = 8 \pi G_N \bar{T}_{\mu\nu} \; , \quad \bar{T}_{\mu\nu} = T_{\mu\nu} - \frac{1}{2} g_{\mu\nu} T
\ee
where $\bar{T}_{\mu\nu}$ is the trace reversed stress tensor. Then the principal part of the left-hand side is $- \frac{1}{2} g^{\alpha\beta} \partial_\alpha \partial_\beta g_{\mu\nu}$, and is hyperbolic in character with each metric component propagating with a light-cone determined by the inverse metric $g^{\mu\nu}$. The contracted Bianchi identity together with stress-energy conservation implies,
\be
\nabla^2 v_\mu + R_\mu^{~\nu} v_\nu = 0
\ee
and so the vector field $v^\mu$ also obeys a wave-like equation. The physical solutions are those with $v_\mu = 0$, as then the harmonic Einstein equation coincides with the actual Einstein equation. Hence if we evolve initial data where $v_\mu = 0$ and $\dot{v}_\mu = 0$, the above equation implies $v_\mu$ will vanish for future times, and the solution of the harmonic Einstein equation will also be a solution of the Einstein equation, now in a coordinate system determined by the condition that $v_{\mu}$ vanishes, which may be viewed as a generalized harmonic gauge condition. The conditions that $v_\mu = 0$ can be viewed as a coordinate restriction on the initial data. From the harmonic Einstein equation one finds that if $v_\mu$ is chosen to vanish on the initial data surface, then the condition $\dot{v}_{\mu} = 0$  simply yields the Hamiltonian and momentum constraints.

\subsection{Harmonic dRGT formulation}
\label{sec:harmform}

This GR harmonic formulation arises due to the introduction of a reference metric. One might wonder whether dRGT, which necessarily has a reference metric, admits a similar formulation. For the minimal theory that we focus on here a very similar formulation exists. We may write $\mathcal{E}^{H}_{\mu\nu}$ as the harmonic version of the (trace reversal of the) Einstein equation in~\eqref{eq:dRGTEinsteinEq} where,
\be
\mathcal{E}^{H}_{\mu\nu} \equiv R_{\mu\nu} - 2\nabla_{(\mu} \xi_{\nu)}  + m^2 \bar{M}_{\mu\nu}  - 8 \pi G_N \bar{T}_{\mu\nu} = 0
\ee
with $\bar{M}_{\mu\nu}$ being the trace reversed mass term,
\be
\bar{M}_{\mu\nu}   = {M}_{\mu\nu}  - \frac{1}{2} M^\alpha_{~\alpha} g_{\mu\nu} = - E_{\mu\nu} - \frac{1}{2} E^\alpha_{~\alpha} g_{\mu\nu} + 3 g_{\mu\nu} 
\ee
and the vector field $\xi^\mu$ is that defined from the vector constraint in~\eqref{eq:xieq}. Taking the divergence of this equation we find,
\be
\label{eq:harmonicBianchi}
\nabla^2 \xi_\mu + R_\mu^{~\alpha} \xi_\alpha =  m^2 \eta_{\mu \alpha} (E^{-1})^{\alpha\beta} \xi_\beta
\ee
again using the contracted Bianchi identity and stress energy conservation. Thus as with the GR harmonic formulation,  we see that if  the initial data obeys $\xi_\mu = 0$ and $\dot{\xi}_{\mu} = 0$, then the vector constraint $\xi_\mu$ will remain zero for all later times. Similarly to the GR case, the condition $\dot{\xi}_\mu = 0$ on the initial data is again equivalent to the Hamiltonian and momentum constraints of the original theory, $\mathcal{E}^t_{~\mu} = 0$, provided the constraint $\xi_\mu = 0$ also vanishes on the initial data surface. 

The scalar constraint now simply follows from the trace of the harmonic Einstein equation, with $\mathcal{S} \equiv - g^{\mu\nu} \mathcal{E}^H_{\mu\nu} = 0$ giving the previous equation~\eqref{eq:scalarconstraint}. 
In the previous formulation of dRGT dynamics above we had 5 components of the vierbein propagate with second order dynamics, with 4 of the 10 vierbein components being removed by the vector constraint and one by the scalar constraint. In this harmonic formulation we no longer have a vector constraint, so that now we expect 9 components will propagate with second order dynamics, and one linear combination does not due to the still present scalar constraint.
About a solution to dRGT with $\xi_\mu = 0$ we can consider fluctuations, and then we may  view these 9 degrees of freedom as the 5 physical graviton modes which preserve $\xi_\mu = 0$, together with 4 constraint violating modes which have $\xi_\mu \ne 0$. The dynamics of $\xi_\mu$ for these constraint violating modes is then governed by equation~\eqref{eq:harmonicBianchi}, and of particular relevance for later is that the principal part of this is, $\sim g^{\alpha\beta} \partial_\alpha \partial_\beta \xi_\mu$, so these four constraint violating modes should be governed by the lightcone of the usual inverse metric.

We may conveniently write the harmonic Einstein equations in terms of the symmetric vierbein by introducing its antisymmetric derivative,
\be
K_{\mu\nu\alpha} = \partial_\mu E_{\nu\alpha} - \partial_\nu E_{\mu\alpha} \; .
\ee
Then we find the harmonic Ricci tensor can be written as\footnote{Note that this choice of coefficients $\mathcal{B}$ and $\mathcal{C}$ is not unique, as can be seen from the identity $K_{\mu\nu\alpha} = 2\left(\partial_{(\mu}E_{\alpha)\nu} - \partial_{(\nu}E_{\alpha)\mu}\right).$},
\be
\label{eq:harmK}
\mathcal{E}^{H}_{\mu\nu} &=& \mathcal{A}^{\sigma\alpha\rho\beta}_{\mu\nu} \partial_\sigma K_{\alpha\rho\beta} + \left( \mathcal{B}^{\sigma\rho\delta\alpha\beta\gamma}_{\mu\nu} +  \mathcal{C}^{\sigma\rho\delta[\alpha\beta]\gamma}_{\mu\nu} \right) K_{\sigma\rho\delta} K_{\alpha\beta\gamma} + \mathcal{C}^{\sigma\rho\delta(\alpha\beta)\gamma}_{\mu\nu} K_{\sigma\rho\delta} \partial_{(\alpha} E_{\beta)\gamma} \nl
&& \qquad  + m^2 \bar{M}_{\mu\nu}  - 8 \pi G \bar{T}_{\mu\nu} 
\ee
with,
\be
\mathcal{A}^{\sigma\alpha\rho\beta}_{\mu\nu} &=& (E^{-1})^{\rho\beta} \delta^\sigma_{(\mu} \delta^\alpha_{\nu)} + g^{\sigma\rho} \delta^\alpha_{(\mu} (E^{-1})^\beta_{\nu)} \nl
\mathcal{B}^{\sigma\rho\delta\alpha\beta\gamma}_{\mu\nu} & = &
 g^{\rho\beta}  \left(
  - \frac{1}{2} \eta^{\delta\gamma} \delta^\sigma_{(\mu} \delta^\alpha_{\nu)} + \frac{1}{4} (E^{-1})^\delta_{(\mu} (E^{-1})^\gamma_{\nu)}g^{\alpha\sigma} + (E^{-1})^{\alpha\gamma} (E^{-1})^\delta_{(\mu} \delta^\sigma_{\nu)}
\right) \nl
\mathcal{C}^{\sigma\rho\delta\alpha\beta\gamma}_{\mu\nu} & = & (E^{-1})^{\gamma\rho} (E^{-1})^\delta_{~(\mu} \left( \delta^\beta_{\nu)} g^{\sigma\alpha} - \delta^\sigma_{\nu)} g^{\alpha\beta}  \right) \nl
&& + (E^{-1})^{\rho\beta}  \left(
  (E^{-1})^{\sigma\delta}  \delta^\alpha_{(\mu} \delta^\gamma_{\nu)} - \frac{3}{2} (E^{-1})^{\gamma\delta}  \delta^\alpha_{(\mu} \delta^\sigma_{\nu)} + \frac{1}{2} (E^{-1})^{\delta\alpha}  \delta^\gamma_{(\mu} \delta^\sigma_{\nu)} 
  \right) \; .
\ee
An interesting point is that we have been able to chose to arrange the terms so that there are no quadratic derivatives terms in $\partial_{\alpha} E_{\beta\gamma}$ -- the quadratic first derivative terms may all be packaged in the antisymmetrized first derivative combination $K_{\mu\nu\alpha}$. The reason for this elegant form can be traced back to the action, which takes the form~\cite{deRham:2023ngf},
\be
S = \frac{1}{8 \pi G_N} \int d^4x \left| \det{E} \right| \left( - \frac{1}{2} A^{\alpha\beta\gamma\mu\nu\sigma} \partial_{[\alpha} E_{\beta]\gamma} \partial_{[\mu} E_{\nu]\sigma} - m^2 \left( E^\alpha_{~\alpha} - 3 \right)  \right) + S_{matter}
\ee
where the tensor $A^{\alpha\beta\gamma\mu\nu\sigma} $ is as given in~\eqref{eq:Atensor}, and we recall it depends on the vierbein and reference metric (but not their derivatives). The Einstein equations come from the equations of motion for the vierbein. 
Varying with respect to the components $E_{\mu\nu}$ then the form of the action together with the fact that $K_{\mu\nu\alpha} = 2 \partial_{[\mu} E_{\nu]\alpha}$ implies that  the two-derivative terms will be packaged in terms of derivatives of $K_{\mu\nu\alpha}$. Further it implies  that quadratic one-derivative terms are given entirely in terms of $ K_{\mu\nu\alpha}$, as above in~\eqref{eq:harmK}.

We again introduce a $3+1$ split, taking the same momentum variables $P_i$ and $P_{ij}$ as before, where these are simply given in terms of the $K_{t\mu\nu}$ components of $K_{\alpha\mu\nu}$ as,
\be
P_i = K_{tit} \; , \quad P_{ij} = K_{tij} \; .
\ee 
For all 9 components of the vierbein except $E_{tt}$, so $E_{ti}$ and $E_{ij}$, to have second order dynamics, then we must have evolution equations for all the momenta $P_i$ and $P_{ij}$ that determine their time derivatives. As before, we may then algebraically determine $E_{tt}$ from the scalar constraint, which contains no derivatives of $E_{tt}$ when written in terms of the vierbein and these momenta.
Naively  we might think that we could then determine $\dot{P}_i$ and $\dot{P}_{ij}$ and $E_{tt}$ from the harmonic Einstein equations $\mathcal{E}^{H}_{\mu\nu}$; $E_{tt}$ being algebraically determined from the trace, and $\dot{P}_i$ and $\dot{P}_{ij}$ from the remaining 9 linearly orthogonal combinations of components.
However it is not quite this simple, as while the trace does determine $E_{tt}$ algebraically, there are time derivatives of $E_{tt}$ in the other components of the harmonic Einstein tensor.
Explicitly the third term above in~\eqref{eq:harmK} does contain  derivatives of $E_{tt}$ as,
\be
\mathcal{E}^{H}_{\mu\nu} &=& \ldots + C^{\sigma\rho\alpha\delta t t}_{\mu\nu} K_{\sigma\rho\alpha} \partial_\delta E_{tt}  + \ldots
\ee
and these include time derivatives.
Thus to close the dynamical system we must determine $\partial_t E_{tt}$ which we may find by taking the time derivative of the scalar constraint $\mathcal{S} = g^{\mu\nu} \mathcal{E}^H_{\mu\nu}$, where,
\be
\label{eq:dtS}
0 = \partial_t \mathcal{S} = \mathcal{S}_1 \partial_t E_{tt} + 2 A^{\alpha\beta\gamma\mu\nu\rho} K_{\alpha\beta\gamma} \partial_t K_{\mu\nu\rho} + \mathcal{S}_2
\ee
where $\mathcal{S}_{1,2}$ are (complicated) coefficients given in terms of $E_{tt}$ (with no derivatives) and the other vierbein components and their first derivatives. Noting that $\mathcal{S}_1$ doesn't vanish near flat spacetime, this then allows us to eliminate the time derivative of $E_{tt}$ in terms of the momenta, and first and second derivatives of the other vierbein components.

We then have two options; 
\begin{enumerate}
\item Determine $\dot{P}_i$ and $\dot{P}_{ij}$ from 9 components of the harmonic Einstein condition and evolve $E_{tt}$ via the time derivative of the scalar constraint, but \emph{not} impose the scalar constraint directly -- the initial data must be chosen so that $\mathcal{S}$ vanishes initially, and since we impose $0 = \partial_t \mathcal{S}$ then it will remain zero.
\item Eliminate $E_{tt}$ and its time derivative entirely from the harmonic Einstein equations using the scalar constraint together with $0 = \partial_t \mathcal{S}$. Then evolve $\dot{P}_i$ and $\dot{P}_{ij}$ using the traceless part of the harmonic Einstein equation.
\end{enumerate}

An important point is that since $\partial_t \mathcal{S}$ involves time derivatives of momenta, through the second term on the r.h.s. of~\eqref{eq:dtS}, and the traceless harmonic Einstein equation contains $\partial_t E_{tt}$, then for either option, these equations form a coupled linear system which determines $\dot{P}_i$, $\dot{P}_{ij}$ and $\partial_t E_{tt}$.
While naively one might think $\dot{P}_i$, $\dot{P}_{ij}$ are determined relatively straightforwardly by the traceless harmonic Einstein equations, this coupling to $\partial_t \mathcal{S}$ makes this very much more complicated. 
\footnote{For linear perturbations about flat spacetime this is much simpler, as the term involving $\partial_t K_{\mu\nu\rho}$ in $\partial_t \mathcal{S}$ vanishes at linear order so $\partial_t \mathcal{S}$ can be solved first for $\partial_t E_{tt}$, and then $\dot{P}_i$ and $\dot{P}_{ij}$ are determined from the traceless $\mathcal{E}^{H}_{\mu\nu}$. }

Precisely the same can be said about the characteristics of the system. Naively the principal part of the harmonic Einstein equation is $\mathcal{E}^{H}_{\mu\nu} {=}_{PP} A^{\sigma\alpha\rho\beta}_{\mu\nu} \partial_\sigma K_{\alpha\rho\beta}$, which is relatively simple. However due to the presence of the term $\partial_\alpha E_{tt}$ in these equations, they must be eliminated by taking derivatives of the scalar constraint, which then also involves terms with $\partial_\sigma K_{\alpha\rho\beta}$ with complicated coefficients.

In order to give a first order formulation, the equations we will now focus on are the 9 components of the harmonic Einstein equation, $\mathcal{E}^{H}_{t i}$ and $\mathcal{E}^{H}_{ij}$, together with the time derivative of the scalar constraint,  $\partial_t \mathcal{S}$, and so we will evolve option 1 above.
We will now specialize to the case of the vacuum equations, setting the stress tensor to vanish. We expect that if a well-posed vacuum formulation can be found, then it will remain well-posed when coupled to reasonable matter.\footnote{
By reasonable matter we mean matter that can be coupled to GR to give a well-posed system.
}

\section{Review of first order well-posedness}
\label{sec:reviewwellposed}

Before we proceed it is useful to review the first order form for a pde system with wavelike behaviour together with the notion of well-posedness. Well-posedness is a crucial feature of any dynamical system of equations that guarantees there is a continuous map from initial data to the subsequent time evolved solutions. While for ill-posed systems one may be able to find particular solutions, perhaps even analytically, without this map from initial data to the space of solutions one cannot regard such a theory as a \emph{dynamical} system that is specified by initial data, and obviously this is the crucial feature we require for our physical theories.

One approach to showing well-posedness is to show that the system can be cast into a `strongly hyperbolic' first order form although we note that lack of such a form doesn't prove the system is ill-posed, as well-posedness depends critically on the way a system is formulated -- there may exist other formulations which are well-posed.
An excellent discussion on `strong hyperbolicity'  is given in ~\cite{Papallo:2017qvl} and we briefly review the relevant aspects here, giving some toy examples.

Let us consider a dynamical p.d.e. system that can be written in first order form in terms of a vector of fields $\matvec{u}(t,\vec{x})$. Consider the linearization of this about a solution $\matvec{\bar{u}}$,
\be
\matvec{u} = \matvec{\bar{u}} + \epsilon \, \delta \matvec{u}
\ee
with $\epsilon \to 0$, is given by the equation,
\be
\label{eq:linsys}
\mat{A}[ \bar{u} ] \cdot ( \partial_t \delta \matvec{u}  ) + \mat{P}^i[ \bar{u} ] \cdot ( \partial_i \delta \matvec{u}  ) + \mat{C}[ \bar{u} ] \cdot \delta  \matvec{u} + \matvec{S}[ \bar{u} ]= 0
\ee
where $\mat{A}$, $\mat{P}^i$, $\mat{C}$ are functions of $t$ and $\vec{x}$ and the background solution $\matvec{\bar{u}}$. In order to solve for $\partial_t \delta \matvec{u}$ then $\mat{A}$ must be invertible. Given a spatial wavevector $k_i$,  we construct from this the matrix,
\be
\mat{M}({k}_i) \equiv -\mat{A}^{-1} \mat{P}^i {k}_i 
\ee
and this encodes the principal symbol of the system.
Roughly speaking this controls the time evolution `in the direction' of the wavevector $k_i$ at a spacetime point.
Then a sufficient condition for the system to be well-posed is that it is strongly hyperbolic:
\footnote{
We emphasize that this is a sufficient but not necessary condition. A weaker requirement is the existence of a symmetriser as discussed in \cite{Sarbach:2012}.
}
\\

\noindent
{\bf Def:} 
The linear pde system in equation~\eqref{eq:linsys} is \textit{strongly hyperbolic}
about the background $\bar{u}$
if for any unit wavevector $\hat{k}_i$ (so $\hat{k}_i \hat{k}_i = 1$) and spacetime point, the matrix $\mat{M}(\hat{k}_i)$ is diagonalizable with real eigenvalues that vary smoothly in spacetime.
\\

The proof of this is given in ~\cite{taylor2010partial} and an excellent review can be found in ~\cite{Papallo:2017qvl, Sarbach:2012}.
We may motivate this as follows. 
Firstly it is the linearization of the p.d.e. system that controls well-posedness, as the question of well-posedness concerns the continuity of the map between initial data and solutions: if we deform the initial data for some solution is there a corresponding deformation to a unique new solution? A necessary condition is that there is a linear map from the tangent space of the space of initial data at some point, to the tangent space of the space of solutions at the corresponding solution. Hence this is determined by the linearization of the p.d.e. system.
If there were no spatial dependence, and hence no spatial derivatives, such a dynamical system would simply be a coupled system of ordinary differential equations, and one could guarantee the (short-time) existence of a unique solution from initial data, provided the coefficient functions were suitably smooth. However for spatially varying fields this may fail due to the short-distance behaviour of the system. Focussing on short distance behaviour, the terms that dominate are those involving derivatives, and so we may neglect the terms with $ \mat{C}$ and $ \matvec{S}$. Further consider the system in the neighbourhood of some spacetime point $(t_{(0)}, x_{(0)}^i)$. 
For solutions that vary on very short wavelengths, and consequently on very short timescales, the slow variation of  $\mat{A}$ and $ \mat{P}^i$ is irrelevant and we may consider these to be `frozen', so constant. Then a formal solution can be given by the Fourier transform as,
\be
\label{eq:formalsoln}
\delta \matvec{u}(t,x^i) = \int d\vec{k} e^{- i k_i (x^i - x^i_{(0)})} \exp( i \mat{M}(k_i) (t - t_0) ) \cdot \matvec{v}(k_i)  \nl
\ee
where $\matvec{v}(k_i)$ give the initial data at the time $t = t_0$ in the spatial neighbourhood of $x^i_{(0)}$.
The question then is: is there a linear map from initial data $\matvec{v}(k_i)$ to the solution $\delta \matvec{u}(t,x^i)$, without placing unreasonable restrictions on the initial data?
Clearly this is true if the Fourier integral  in the above formal solution converges. A sufficient condition is that there exists a function of time, $f(t)$, such that for any $k_i$ and time $t > t_0$,
\be
\label{eq:normcondition}
\left\| \exp( i \mat{M}(k_i) (t - t_0) ) \right\| \le f(t - t_0)
\ee
which further implies the norm of the solution at time $t$ is bounded by that at $t_0$, so $\left\| \delta u \right\|(t) = f(t - t_0) \left\| \delta u \right\|(t_0)$, for any choice of norm. This then implies well-posedness.
If $\mat{M}(\hat{k}_i)$ is diagonalizable with real eigenvalues for any unit $\hat{k}_i$, so that $\hat{k}_i \hat{k}_i = 1$, then~\eqref{eq:normcondition} is guaranteed, hence the definition of strong hyperbolicity above.
 One might think that simply having real eigenvalues is sufficient as a condition, so called weak hyperbolicity. However it is not the case -- the matrix is also required to have non-trivial eigenvectors, as we see in the examples that we give shortly.

Suppose the system is well-posed. What then is the physical interpretation of the eigenvectors? Suppose at a spacetime point $p$ one such eigenvector is $\matvec{v}(\hat{k}_i)$, so that,
\be
\mat{M}(\hat{k}_i) \cdot \matvec{v}(\hat{k}_i) = \Lambda(\hat{k}_i) \matvec{v}(\hat{k}_i)
\ee
where $\Lambda(\hat{k}_i)$ is then the corresponding eigenvalue given the unit wavevector $\hat{k}_i$. We may interpret this as giving a solution to the linearized system in the short wavelength -- i.e. geometric optics -- limit. 
Then locally near $p$ a solution $\delta \matvec{u}$ of the equations linearized about the background $\matvec{\bar{u}}$ is,
\be
\delta \matvec{u} \simeq \epsilon \matvec{v}(\hat{k}_i) e^{-i \zeta \hat{k}_\mu x^\mu} \; , \quad \hat{k}_\mu = \left( + \Lambda(\hat{k}_i), \hat{k}_i \right)
\ee
in the limit that $\zeta \to \infty$. This goes as $\sim e^{-i {k}_\mu x^\mu}$ where $k_\mu = \zeta \hat{k}_\mu$ is the physical wavevector which is very large giving the geometric optics limit.
The spacetime wavevector $\hat{k}_\mu$ then gives one of the characteristic directions of the system at the point $p$.

\subsection{Some examples}
\label{sec:example}

To be more concrete, let us illustrate this with the example of a d-dimensional linear wave equation for a field $\phi(t,\vec{x})$ given as,
\be
g^{\mu\nu} \partial_\mu \partial_\nu \phi + v^\mu \partial_\mu \phi + V \phi +S = 0
\ee
controlled by the inverse metric $g^{\mu\nu}(t,\vec{x})$, vector field $v^{\mu}(t,\vec{x})$, non-linear potential $V(t,\vec{x})$ and source function $S(t,\vec{x})$. 
We expect this should be a well-posed wavelike theory provided that $g^{\mu\nu}$ is the inverse of a Lorentzian metric. We may write this in first order form by introducing new variables, $\pi_\mu = \partial_\mu \phi$ and then taking,
\be
\matvec{u} = \left( \pi_t, \pi_i , \phi \right)
\ee
so that in first order form we have,
\be
\mat{A} = \left( 
\begin{array}{ccc}
g^{tt} &  g^{tj} & 0 \\
0 & \delta^{ij} & 0 \\
0 & 0 & 1
\end{array}
\right) \; , \quad 
\mat{P}^k = \left( 
\begin{array}{ccc}
g^{tk} & g^{kj} & 0 \\
- \delta^{ik} & 0 & 0 \\
0 & 0 & 0
\end{array}
\right) \nonumber
\ee
\be
\mat{C} =  \left( 
\begin{array}{ccc}
v^{t} & v^{j} & V \\
0 & 0 & 0 \\
1 & 0 & 0
\end{array}
\right) \; , \quad 
\matvec{S} = \left( 
\begin{array}{c}
S \\
0  \\
0 
\end{array}
\right)
\ee
and we have expanded our system, and introduced the new dynamical equations,
\be
\dot{\phi} = \pi_t \; , \quad \dot{\pi}_i = \partial_i \pi_t \; .
\ee
Solutions of this expanded system coincide with those of the wave-equation provided the initial data satisfies the consistency relation $\pi_i = \partial_i \phi$.
The determinant of $\mat{A}$ is $\det(\mat{A}) = g^{tt}$, and hence it is invertible unless $g^{tt}$ vanishes. This is equivalent to the condition that $g^{\mu\nu} n_\mu n_\nu \ne 0$ where $n_\mu = (1,0,\ldots,0)$ so that $n = dt$. Hence we may invert $\mat{A}$ provided the constant $t$ surfaces are nowhere null with respect to the metric $g_{\mu\nu}$.
Now,
\be
\mat{M}(k_i) \equiv - \mat{A}^{-1} \mat{P}^i k_i 
\ee
and this has the two physical eigenvalues, $\Lambda_{\pm}$ which are solutions of the quadratic light cone condition,
\be
g^{\mu\nu} k_\mu k_\nu = 0 \; , \quad k_\mu = \left( +\Lambda, k_i \right)
\ee
so that the associated eigenvectors are $\left( \Lambda, k_i , 0 \right)$,
and $d-1$ auxiliary zero eigenvalues associated to writing the system in first order form, whose eigenvectors are,
\be
\left( 0, \vec{0}, 1 \right) \; , \quad \left( 0, u^{(a)}_i , 0 \right)
\ee
for $a = 1,\ldots,d-2$, where the non-zero components of $u^{(a)}_i$ are,
$u^{(a)}_{1} = - g^{a i} k_i$ and $u^{(a)}_{a+1} = g^{1 i} k_i$.
Thus the system is well-posed provided the metric $g_{\mu\nu}$ is Lorentzian, which ensures that the two eigenvalues $\Lambda_{\pm}$ are distinct and real.

It is instructive to consider some ill-posed examples. Perhaps the most obvious one is to try to solve an elliptic p.d.e. by a Cauchy evolution. An example of this is to take the equation above with Riemannian metric $g_{\mu\nu}$, in which case one does not have a lightcone structure, and correspondingly the eigenvalues $\Lambda_{\pm}$ will be imaginary, violating the well-posedness condition.

A more subtle example can be given taking a simple case of two uncoupled wave equations in 2d with a flat metric, so $\ddot{\phi} - \partial_x^2 \phi = 0$ and $\ddot{\psi} - \partial_x^2 \psi = 0$. From above we would define $\matvec{u} = \left( \pi_t = \dot{\phi}, \pi_x = \phi' , \phi , \Pi_t = \dot{\psi}, \Pi_x = \psi', \psi \right)$ and obtain the matrices,
\be
\mat{M}(\hat{k}_x) =  \left( 
\begin{array}{cccccc}
0 & 1 & 0 & 0 & 0 & 0  \\
1 & 0 & 0 & 0 & 0 & 0\\
0 & 0 & 0 & 0 & 0 & 0 \\
0 & 0 & 0 & 0 & 1 & 0  \\
0 & 0 & 0 & 1 & 0 & 0  \\
0 & 0 & 0 & 0 & 0 & 0  \\
\end{array}
\right)
 \quad \implies \quad 
 \mathrm{Jordan}\left[ \mat{M} \right] = \left( 
\begin{array}{cc|cc|cc}
1 & 0 & 0 & 0 & 0 & 0  \\
0 & 1 & 0 & 0 & 0 & 0 \\ \hline
0 & 0 & -1 & 0 & 0 & 0 \\
0 & 0 & 0 & -1 & 0 & 0  \\ \hline
0 & 0 & 0 & 0 & 0 & 0  \\
0 & 0 & 0 & 0 & 0 & 0  \\
\end{array}
\right) 
\ee
with its diagonal Jordan form, showing the expected physical eigenvalues $\pm 1$ twice, once for each field, and two zero eigenvalues associated to the derivative variables that have been introduced.

However we might instead make the choice to extend our variables as $\pi_t = \dot{\phi}$, $\pi_x = \phi'$ as before but now taking, $\Pi_t = \dot{\psi} - \phi'$, $\Pi_x = \psi'$. Then the matrix becomes,
\be
\mat{M}(\hat{k}_x) =  \left( 
\begin{array}{cccccc}
0 & 1 & 0 & 0 & 0 & 0  \\
1 & 0 & 0 & 0 & 0 & 0\\
0 & 0 & 0 & 0 & 0 & 0 \\
-1 & 0 & 0 & 0 & 1 & 0  \\
0 & 1 & 0 & 1 & 0 & 0  \\
0 & 0 & 1 & 0 & 0 & 0  \\
\end{array}
\right)
\quad \implies \quad 
 \mathrm{Jordan}\left[ \mat{M} \right] = \left( 
\begin{array}{cc|cc|cc}
1 & 0 & 0 & 0 & 0 & 0  \\
0 & 1 & 0 & 0 & 0 & 0\\ \hline
0 & 0 & -1 & 0 & 0 & 0 \\
0 & 0 & 0 & -1 & 0 & 0  \\ \hline
0 & 0 & 0 & 0 & 0 & 1  \\
0 & 0 & 0 & 0 & 0 & 0  \\
\end{array}
\right) 
\ee
and we see that the Jordan block associated to the zero eigenvalue becomes non-diagonal; thus there is only one eigenvector with zero eigenvalue, and the matrix $\mat{M}$ does not have a complete set of eigenvectors. We note that the system still has the two sets of eigenvalues $\pm 1$ associated to the physical propagation, but the ill-posedness has arisen from the way the system has been extended to a first order system. We also note that the eigenvalue of the eigenvectors that do exist are real here -- so this example is weakly hyperbolic, but fails to be strongly hyperbolic.

It is instructive to see how the problem arises in this example. The general solution in both cases can be given in terms of the spatially Fourier transform of the fields. Writing ${\phi}(t,x) =\frac{1}{2\pi} \int dk e^{-i k x} \tilde{\phi}(t,k)$, and similarly for the other fields, the general solution in the first (well-posed) case is,
\be
\begin{array}{rclrcl}
\tilde{\pi}_t &= & k \left( - \sin(k t) \left( c_\phi +  \phi_0\right) + \cos(k t) \pi_0 \right) &
\tilde{\Pi}_t &= & k \left( - \sin(k t) \left( c_\psi +  \psi_0\right) + \cos(k t) \Pi_0 \right) \nl
\tilde{\pi}_x &= & - i k \left( \cos(k t) \left( c_\phi +  \phi_0\right) + \sin(k t) \pi_0 \right) &
\tilde{\Pi}_x &= & - i k \left( \cos(k t) \left( c_\psi +  \psi_0\right) + \sin(k t) \Pi_0 \right) \nl
\tilde{\phi} & = & \phi_0 +  \sin(k t) \pi_0 + \left( \cos(k t)  - 1 \right) \left( c_\phi +  \phi_0\right)  &
\tilde{\psi} & = & \psi_0 + \sin(k t) \Pi_0 + \left( \cos(k t)  - 1 \right) \left( c_\psi +  \psi_0\right) 
\end{array}
\ee
where we have 6 constants of integration associated to the 6 first order time integrations, namely $\phi_0$, $\pi_0$, $c_\phi$, $\psi_0$, $\Pi_0$, $c_\psi$. It is important that we have introduced these so that in the large $k$ limit, the terms controlled by each of these in each of the variables are equally important.
Then we have that at time $t = 0$,
\be
\left. \tilde{\phi} \right|_{t=0} = \phi_0 \; , \quad \left. \frac{d}{dt} \tilde{\phi} \right|_{t=0} = k \pi_0  \; , \quad  \left. \tilde{\psi} \right|_{t=0} = \psi_0 \; , \quad \left. \frac{d}{dt} \tilde{\psi} \right|_{t=0} = k \Pi_0 
\ee
so that $\phi_0$, $\pi_0$, $\psi_0$, $\Pi_0$ give the `physical' data for the wave equations. The remaining two constants of integration, $c_{\phi}$ and $c_{\psi}$, are due to the extension of the system when formulated in first order form. For a general solution they are free, but if we require the solution to be that of the two wave-equations then we must ensure the conditions $\pi_x = \partial_x \phi$ and $\Pi_x = \partial_x \psi$ apply to the initial data; here these simply imply $c_\phi = 0$ and $c_\psi = 0$. 

On the other hand for the second formulation the general solution is the same for $\tilde{\pi}$ and $\tilde{\Pi}_{t,x}$ but now,
\be
\begin{array}{rclrcl}
\tilde{\Pi}_t &= & k \left( - \sin(k t) \left( c'_\psi +  \psi_0 - i \pi_0 \right) + \cos(k t) \left( \Pi'_0 + i c_\phi + i  \phi_0 \right) \right) \nl
\tilde{\Pi}_x &= & - i k \left( \cos(k t) \left( c'_\psi + \psi_0\right) + \sin(k t) \Pi'_0 \right) \nl
\tilde{\psi} & = & \psi_0 + i k t c_\phi + \sin(k t) \Pi'_0 + \left( \cos(k t)  - 1 \right) \left( c'_\psi + \psi_0 \right)
\end{array}
\ee
where the constants of integration are now $\phi_0$, $\pi_0$, $c_\phi$, $\psi_0$, $\Pi'_0$, $c'_\psi$. At time $t = 0$ we obtain the same values for $\tilde{\phi}$ and its time derivative, and also for $\tilde{\psi}$ as above, but now the time derivative of $\tilde{\psi}$ differs as,
\be
\label{eq:timederiv}
 \left. \frac{d}{dt} \tilde{\psi} \right|_{t=0} = k \left( \Pi'_0 + i c_{\phi} \right) \; .
\ee
Now requiring the solution of this expanded first order system to be consistent with a solution to the two wave equations implies we constrain the initial data, again imposing $\pi_x = \partial_x \phi$ and $\Pi_x = \partial_x \psi$, and these conditions imply that the constants $c_\phi$ and $c'_{\psi}$ vanish.

We immediately see that for solutions to the original wave equation problem, these two formulations agree; $c_\phi = c_\psi = c'_\psi = 0$ and $\Pi'_0 = \Pi_0$, and the solution for the original fields, $\tilde{\phi}$ and $\tilde{\psi}$ are identical as they should be. The issue is that the general solutions of these two formulations differ -- and we see explicitly that in the second ill-posed formulation there is a linear growth in $\tilde{\psi}$ in the case that $c_\phi \ne 0$ going as $\sim k t$. 
We can then construct sequences of solutions for increasing $k$ such that as $k \to \infty$ these solutions would have the same initial data, but different time evolutions.
This example shows explicitly that weak hyperbolicity is not sufficient for well-posedness.

If we were to use these first order formulations to perform numerical evolutions, numerical error would inevitably lead to an initial non-zero $c_\psi$, and then for very large $k$ these would quickly grow to be appreciable and destabilize the solution from the one of physical interest.\footnote{
One might naively think that we could rescale $c_\phi \to \frac{1}{k} c_\phi$ and then this linear growth would only go as $\sim t$ rather than as $\sim k t$ and not be a problem. However it is important that these constants of integration were chosen so that they all enter the other physical variables  with they same scaling as $k \to \infty$. Hence this rescaling would amount to setting $c_\phi$ to zero in this large $k$ limit, and represents a fine tuning of the initial data which is precisely what we try to avoid in using a well-posed system.
}

\section{A naive first order form for dRGT}
\label{sec:naive}

We will now consider dRGT with the view to obtaining a well-posed first order formulation. For simplicity we will consider vacuum, so $T_{\mu\nu} = 0$. We expect that, as with GR, the addition of reasonable matter that  couples minimally to the metric and its connection will not change the issue of well-posedness.

Following the discussion above we take variables for dRGT in first order form to be the symmetric vierbein components, $E_{\mu\nu}$, which we write as,
\be
E_{\mu\nu} = \left( 
\begin{array}{cc}
\phi & V_i \\
& e_{ij} 
\end{array}
\right)
\ee
together with the momenta $P_i$ and $P_{ij}$. However looking at the example of the wave-equation above, we also require new variables for spatial derivatives of fields, and these we will take to be the following components of $K_{\mu\nu\alpha}$,
\be
Q_{ij\mu} \equiv K_{ij\mu} = \partial_i E_{j\mu} - \partial_j E_{i\mu} \; .
\ee
These will expand our system; they will also imply that initial data should be chosen consistent with the above in order for it to evolve as a solution of the dRGT theory.
Since $e_{ij}$ is symmetric we choose the variables we work with to be the ones where $ij$ are ordered in an even permutation -- so the diagonal elements, together with $e_{xy}$, $e_{yz}$ and $e_{zx}$. We do this to ensure that rotational symmetry is preserved.
We then take our vector of fields to be,
\be
\matvec{u} = \left(
\phi, P_i, P_{ij}, V_i , e_{ij}, Q_{ij\mu} 
\right)
\ee
where the pair $ij$ is understood to be in $\{ xx, yy, zz, xy, yz, zx \}$. Then there are 3 independent fields and momenta in $V_i$ and $P_i$; 6 independent  fields and momenta in $e_{ij}$ and $P_{ij}$ given the symmetry of these; and 12 independent components of $Q_{ij\mu}$ given its anti-symmetry in its first index pair. Altogether then this yields 31 variables for the first order system.

In terms of evolution equations for the system we have the time derivative of the scalar constraint, $\partial_t \mathcal{S} = 0$, and the harmonic Einstein equation components $\mathcal{E}^{H}_{t i}$ and $\mathcal{E}^{H}_{ij}$. As discussed above we miss the time-time component of the harmonic Einstein equation which is a linear combination of the scalar constraint, $\mathcal{S}$, together with the other components of $\mathcal{E}^{H}_{\mu\nu}$, and which we will not impose as an equation of motion, but rather that its time derivative vanishes (since we are imposing $\partial_t \mathcal{S} = 0$). We may take these collectively to yield evolution equations determining the time derivatives of $\phi$, $P_i$ and $P_{ij}$. From the above equation~\eqref{eq:scalarconstraintexpr} we have,
\be 
\label{eq:dtS2}
\partial_t \mathcal{S} &=& 2 A^{\alpha\beta\gamma\mu\nu\rho} K_{\alpha\beta\gamma} \partial_t K_{\mu\nu\rho} + \left( \partial_t A^{\alpha\beta\gamma\mu\nu\rho} \right) K_{\alpha\beta\gamma} K_{\mu\nu\rho}  + 3 m^2 \partial_t( E^\mu_{~\mu} ) 
\ee
where the first term on the right-hand side involves time derivatives of $P_i$, $P_{ij}$ and $Q_{ij\mu}$, and the last two involve time derivatives of the vierbein components. There are no spatial derivative terms of the fields $\matvec{u}$, and no terms without derivatives of these fields. For the harmonic Einstein equation we have,
\be
\label{eq:Ein2}
\mathcal{E}^{H}_{\mu\nu} &=& A^{\sigma\alpha\rho\beta}_{\mu\nu} \partial_\sigma K_{\alpha\rho\beta} + C^{\sigma\rho\delta\alpha\beta\gamma}_{\mu\nu} K_{\sigma\rho\delta} \partial_{(\alpha} E_{\beta)\gamma} + \ldots
\ee
where $\ldots$ are terms with no derivatives of the fields $\matvec{u}$, and are given explicitly in~\eqref{eq:harmK}. We see the first term on the righthand side involves time and spatial derivatives of $P_i$, $P_{ij}$ and $Q_{ij\mu}$, and the second involves time and spatial derivatives of the vierbein, including the time-time component $\phi$.

Evolution equations for the vierbein variables arise from the definition of the momenta, so,
\be
\dot{V}_i = P_i + \partial_i \phi \; , \quad \dot{e}_{ij} = P_{ij} + \partial_{i} V_{j} 
\ee
and,
\be
\dot{Q}_{ijt} = \partial_i P_{j} - \partial_j P_{i} \; , \quad \dot{Q}_{ijk} = \partial_i P_{jk} - \partial_j P_{ik} 
\ee
where in the second expression we note that if $i \ne j$ and the ordering of  the indices of $P_{ij}$ is opposite to our choice $\{xy, yz, zx \}$, then $P_{ij}$ should be replaced using the identity,
\be
P_{ij} = P_{ji} + Q_{jit}
\ee
which now will be given by our variables.
Putting this together gives the first order system,
\be
\mat{A}[ {u} ] \cdot ( \partial_t \matvec{u}  ) + \mat{P}^i[ {u} ] \cdot ( \partial_i \matvec{u}  ) + \matvec{C}[ \matvec{u} ] = 0
\ee
where we emphasize that this has the nice property that it is linear in derivatives of the fields $\matvec{u}$. 
The $\mat{A}[ {u} ]$, $\mat{P}^i[ {u} ]$ and $\matvec{C}[ \matvec{u} ]$ are functions of spacetime position.
Well-posedness about some field configuration $\bar{\matvec{u}}$ then requires that at all spacetime points,
\be
\mat{M}[\hat{k}_i] = -\mat{A}[ \bar{u} ]^{-1} \cdot \mat{P}^i[ \bar{u} ] \hat{k}_i
\ee
has a complete set of eigenvectors, each with real eigenvalues, for all unit wavevectors $\hat{k}_i$, so that $\hat{k}_i \hat{k}_i = 1$. 
It further requires that these eigenvalues vary smoothly in spacetime.
Since the theory and this formulation has a rotational invariance, we can without loss of generality choose $\hat{k}_i = (1, 0, 0)$ to study  well-posedness, 
where instead of varying $\hat{k}$ we consider the set of backgrounds $\bar{u}$ related by rotations, rather than the single background $\bar{u}$.
We will use the notation,
\be
\mat{M}_x \equiv \mat{M}[(1,0,0)] \; .
\ee

For this system ill-posedness is  already seen taking $\bar{u}$ to be that of the Minkowski spacetime. Then $\bar{\matvec{u}}$ is given by $\phi = -1$, $V_i = 0$ and $e_{ij} = \delta_{ij}$, with all other fields ($P_i$, $P_{ij}$ and $Q_{ij\mu}$) vanishing as they involve derivatives of the vierbein. Evaluating the matrix $\mat{M}[\hat{k}_i]$ for any unit $\hat{k}_i$ we find it can be bought into the Jordan normal form,
\be
\label{eq:MinkJordan}
\mathrm{Jordan}\left[ \mat{M}[\hat{k}_i]  \right]= \left(
\begin{array}{c|c|c|c|c}
\begin{array}{cc} 0 & 1\\0 & 0 \end{array} 
& 0 & 0 & 0 & 0 
\\ \hline
 &  \begin{array}{cc} 0 & 1\\0 & 0 \end{array} 
 &0 & 0 & 0 
  \\ \hline
 & & + \mat{1}_9 
 & 0 & 0     
 \\ \hline
  & &  &  - \mat{1}_9 
  & 0
  \\ \hline
 &  & &  & \mat{0}_9  \\
\end{array}
\right)
\ee
with three 9$\times$9 blocks given by the identity, minus the identity, and zero, and then two off diagonal 2$\times$2 blocks.
Thus while all the eigenvalues of this matrix are real, it does not have a complete set of eigenvectors, with some associated to zero eigenvalue missing. Hence the system linearized about Minkowski is weakly hyperbolic, but not strongly hyperbolic. 

Let us consider the causes of this non-diagonalizable pair of blocks. One is due to the pair of evolution equations,
\be
\partial_t e_{ij} &=& P_{ij} + \partial_i V_j \nl
\partial_t V_j &=& P_{j} + \partial_j \phi \; .
\ee
Considering $\hat{k}_i = (1,0,0)$, so $x$-derivatives, these equations contain the subsystem,
\be
\partial_t e_{xy}   =  \partial_x V_y + \ldots \; , \quad
\partial_t V_y = \ldots 
\ee
where $\ldots$ represent terms with no time or $x$ derivatives. When we linearize about flat spacetime, there are no other terms involving the derivatives of $e_{xy}$ or $V_y$ -- the first two terms in~\eqref{eq:dtS2} vanish since $K_{\mu\nu\alpha}$ vanishes on the background, and while the last term  involves derivatives of the vierbein as $\partial_t E^\mu_{~\mu}$, it only involves diagonal components when evaluated about Minkowski. The terms in~\eqref{eq:Ein2} involving derivatives of the vierbein similarly vanish due to $K_{\mu\nu\alpha}$ vanishing.
As a consequence, the $x$-derivative structure linearized about Minkowski is block diagonal, with this giving one of the blocks, so that,
\be
\label{eq:block}
\mat{M}_x = \left(
\begin{array}{ccc}
0 & 1 & 0  \\
0 & 0 & 0 \\
0 & 0 & \mat{Q}
\end{array}
\right)
\ee
and this 2 by 2 block cannot be diagonalized -- it has two zero eigenvalues, but only one non-trivial eigenvector $(1,0,\matvec{0})$.

The second block arises for similar reasons, from the 6 evolution equations,
\be
 \partial_t Q_{ijt} = \partial_i P_j - \partial_j P_i \nl
 \partial_t Q_{ijk} = \partial_i P_{jk} - \partial_j P_{ik} \; .
\ee
Recalling that $P_{ij} = P_{ji} + Q_{jit}$ and our variables $P_{ij}$ and $Q_{ij\mu}$ should have their $ij$ indices in cyclic order, we write the evolution equation for the $Q_{ijt}$ and $Q_{ijk}$ by defining,
\be
\label{eq:Fdefn}
F_x &\equiv &  \partial_t Q_{xyt} - \partial_x P_y + \partial_y P_x \nl
F'_x &\equiv &   \partial_t Q_{xyz} - \partial_x P_{yz} + \partial_y \left( P_{zx} + Q_{zxt} \right)
\ee
and then defining $F_{y,z}$ and $F'_{y,z}$ from cyclicly permuting the $x,y,z$ indices for these. Then the evolution equations are given by the set of equations $F_i = 0$ and $F'_i = 0$.
These equations include the subsystem $F_y$, $F'_x$ and $F'_z$, where taking $\hat{k}_i = (1,0,0)$ and focussing on $x$-derivatives, we have,
\be
F'_x + F'_z = \partial_t \left(  Q_{xyz} + Q_{zxy} \right) + \partial_x Q_{yzt}  + \ldots \; , \quad
F_y = \partial_t Q_{yzt} + \ldots 
\ee
where again $\ldots$ are terms without time or $x$ derivatives. Again about Minkowski, this gives a diagonal block of the same form as~\eqref{eq:block} as there are no other derivative terms involving $Q_{xyz}$, $Q_{zxy}$ or $Q_{yzt}$ in any of the other equations.

Before we go on to modify the first order system to yield a well-posed system, we pause briefly to discuss the eigenvectors of $\mat{M}$ and the vector constraint. As discussed above, the eigenvectors of $\mat{M}$ at a spacetime point $p$ locally give short wavelength linear fluctuations at that point. Thus given an eigenvector,
\be
\mat{M}(\hat{k}_i) \cdot \matvec{v} = \Lambda \matvec{v}
\ee
we may define $\hat{k}_\mu = (+\Lambda, \hat{k}_i)$ and write the components as,
\be
\matvec{v} = \left(
\delta \phi, \delta P_i, \delta P_{ij}, \delta V_i , \delta e_{ij}, \delta Q_{ij\mu} 
\right)
\ee
and then these give the linear solution, $\phi \simeq \bar{\phi} + \delta \phi \, e^{- i \zeta \hat{k}_\mu x^\mu}$
 in the short wavelength limit $\zeta \to \infty$, where $\bar{\phi}$ is the background $\phi$, 
 and similarly for the other variables, 
 and this is valid near the point $p$ in the linear approximation. For non-zero $\Lambda$ this should correspond to plane-wave solutions to the linearized harmonic Einstein equation. Hence if the background solution at $p$ has vanishing $\xi_\mu$, we can ask whether these linear wavemodes preserve vanishing $\xi_\mu$ or not, so whether they correspond to short wavelength solutions of linearized dRGT or not. We linearize the vector constraint $\xi_\mu$, so,
\be
\delta \xi_\mu = - 2 \bar{g}^{\mu\alpha} (\bar{E}^{-1})^{\beta\sigma} \partial_{[ \alpha} \delta E_{\beta]\sigma} + \ldots
\ee
where $\ldots$ are terms that involve no derivatives of the perturbation and hence are subdominant in the short-wavelength limit. Dropping these terms with no derivatives, and using $\partial_{[ \alpha} \delta E_{\beta]\sigma} = 2 \delta K_{\alpha\beta\sigma}$, where $\delta K_{t i t} = \delta P_{i}$, $\delta K_{t i j} = \delta P_{ij}$ and $\delta K_{i j \mu} = \delta Q_{i j \mu}$
then we may write,
\be
\label{eq:deltaxi}
\delta \xi_\mu \simeq \mat{\Xi}_\mu \cdot \matvec{v} 
\ee
where $\mat{\Xi}_\mu$ only projects onto the derivative fluctuations in $\matvec{v}$, so $\delta P_{i}$, $\delta P_{ij}$ and $\delta Q_{ij\mu}$.
Then we should see the physical modes of dRGT as those eigenvectors of $\mat{M}$ with non-zero eigenvalue that are annihilated by $\mat{\Xi}$.

If we consider the Minkowski background we may w.l.o.g. consider propagation in the $x$-direction, so taking $\hat{k}_i = (1,0,0)$, and then consider $\mat{M}_x \equiv \mat{M}(\hat{k}_i)$. The 9 physical modes with eigenvalue $\Lambda = \pm 1$ are then given by the eigenvectors $\matvec{v}$ where,
\be
\delta P_{x} &=& \mp \left( \delta P_{xx}+\delta P_{yy}+\delta P_{zz} \right) \; , \quad \delta V_x = \pm \delta \phi \; , \quad \delta e_{xx} = \delta \phi
 \nl
 \delta Q_{xyy} &=& \pm \delta P_{yy}   \; , \quad  \delta Q_{zxx} = \mp \delta P_{zx}
  \; , \quad 
 \delta Q_{xyx} = \delta P_{y} \pm \delta P_{xy}  \; , \quad  \delta Q_{zxz} =  \mp \delta P_{zz}
\nl
  \delta Q_{xyt} &=& \pm \delta P_{y}
   \; , \quad 
  \delta Q_{zxt} = \mp \delta P_{z}
   \; , \quad 
  \delta Q_{xyz} = \pm \delta P_{yz}
    \; , \quad 
  \delta Q_{zxy} = \mp \delta P_{yz}
\ee
with all other components vanishing. Written in this way the data for these 9 eigenvectors comprises $\delta \phi$ and the 6 components of $\delta P_{ij}$ together with $\delta P_y$ and $\delta P_z$.
Now to examine the subspace annihilated by $ \mat{\Xi}_\mu$ we compute,
\be
\mat{\Xi}_\mu \cdot \matvec{v} = \left(  \delta P_{xx}+\delta P_{yy}+\delta P_{zz} , \mp \delta P_{xx}, \mp \delta P_{xy}, \mp \delta P_{xz} \right)
\ee
where we should understand the last component in terms of our variables as $\delta P_{xz} = \delta P_{zx} + \delta Q_{zxt}=-\delta P_z +\delta P_{zx}$ for the vector $\vec{v}$.

Thus the 9 modes comprise 4 constraint violating modes. We may take these to be parameterized by the components of $\mat{\Xi}_\mu \cdot \matvec{v}$ above, while setting $\delta \phi$, $\delta P_y$ and $\delta P_z$ to zero as well as the component $\delta P_{yz}$, and also setting $\delta P_{yy} = \delta P_{zz}$. Considering rotations about the $x$ axis, these then comprise two scalar modes, given by the trace $\delta P_{xx}+\delta P_{yy}+\delta P_{zz}$ and the component $\delta P_{xx}$, together with the vector mode $\delta P_{x I}$, so $I= \{ y,z\}$, in the $y$-$z$ plane.

Then the remaining 5 physical modes of dRGT are those preserving the constraint, so being annihilated by $\mat{\Xi}_\mu \cdot \matvec{v}$, and thus having $\delta P_{xi} = 0$ together with $\delta P_{yy} = - \delta P_{zz}$. These are then parameterized by $\delta \phi$, by $\delta P_I$ and by the traceless part of $\delta P_{IJ}$, which respectively give the scalar mode, the vector and the spin-2 modes under rotations about the $x$-axis.

\section{Well-posed formulation for the Minkowski background}
\label{sec:wellposed}

We have seen that the above naive first order formulation fails to be strongly hyperbolic already on the Minkowski background. Further we have seen that the problem lies in the definition of the momentum and spatial derivative variables and their evolution equations -- not the harmonic Einstein equations or (time derivative of the) scalar constraint. Thus we might hope to modify these definitions to obtain a first order system that is strongly hyperbolic. Indeed this is possible, at least for the Minkowski background, as we will now describe.

Firstly we may relatively straightforwardly modify the evolution equations for $Q_{ij\mu}$ to fix the issue described above. The key point is that from the definition of $Q_{ij\mu}$ we see that there is the consistency relation,
\be
\label{eq:consistency1}
I_x \equiv \partial_z Q_{xyz} + \partial_y Q_{zxz} + \partial_x Q_{yzz}  &=& 0
\ee
and two similar relations, $I_{y,z}$, obtained by cyclicly permuting $x$, $y$ and $z$, and also the relation,
\be
\label{eq:consistency2}
J =  \partial_x Q_{yzt} + \partial_y Q_{zxt} + \partial_z Q_{xyt} & = & 0 \; .
\ee
Having introduced the variables $Q_{ij\mu}$, we have expanded our first order system, but are only interested in solutions which reduce to the harmonic Einstein equations. In particular we must ensure that the initial data obeys these constraints $I_i = J = 0$ so that the solution will indeed be that of the harmonic Einstein equations. The full set of conditions that must be placed on the initial data for it to evolve as a solution to the harmonic Einstein equation will be described in the later section~\ref{sec:initialdata}, as well as how they are  preserved by the evolution equations. We can modify our first order system by adding these constraints -- while this changes the first order system, it will not change the solutions we are interested,  those being the ones where these constraints vanish.

We now consider modifying the evolution equations in equation~\eqref{eq:Fdefn} by defining;
\be
\tilde{F}_x &\equiv & F_x - \mu^2 I_x \nl
\tilde{F}'_x &\equiv & F'_x - J 
\ee
for some real positive constant $\mu > 0$.
Now if we consider the case $\hat{k}_i = (1,0,0)$ and focus on $x$-derivatives, the previous troublesome equations are modified to,
\be
\tilde{F}'_x + \tilde{F}'_z = \partial_t \left(  Q_{xyz} + Q_{zxy} \right) - \partial_x Q_{yzt}  + \ldots \, , \;
\tilde{F}'_y   =  \partial_t Q_{yzx} -  \partial_x Q_{yzt} + \ldots \, , \;
\tilde{F}_y = \partial_t Q_{yzt}  - \mu^2  \partial_x Q_{yzx}  + \ldots 
\ee
where now we see the addition of $J$ has also coupled in the equation $\tilde{E}'_y$ and the variable $Q_{yzx}$, where previously $E'_y$ did not involve the variables $Q_{xyz}$,$Q_{zxy}$ or $Q_{yzt}$. Considering the 3 variables $\matvec{v} = \left( Q_{xyz} + Q_{zxy}, Q_{yzx}, Q_{yzt} \right)$ this gives the system,
\be
0 = \partial_t \matvec{v} + \mat{A} \cdot \partial_x \matvec{v} + \ldots 
\ee
and now,
\be
\mat{A} = \left( \begin{array}{ccc}
0 & 0 & -1 \\
0 & 0 & -1 \\
0 & - \mu^2 & 0
\end{array} \right)
\ee
which has a zero eigenvector $(1,0,0)$ and two eigenvectors $(1, 1, \pm \mu)$ with eigenvalues $\mp \mu$. 

A more subtle modification is required in order to fix the well-posedness of the $V_i$ and $e_{ij}$ evolution equations.
We introduce auxiliary variables associated to the off-diagonal vierbein components, $\tilde{e}_{xy}$, $\tilde{e}_{yz}$ and $\tilde{e}_{zx}$, and we evolve these by the same equations as for the off-diagonal vierbein, ${e}_{xy}$, ${e}_{yz}$ and ${e}_{zx}$, but modified using the following consistency condition,
\be
\label{eq:consistencyQijt}
 Q_{ijt} - \partial_i V_j + \partial_j V_i = 0 \; .
\ee
Again this consistency condition must be imposed on the initial data for the system if the solution is to correctly correspond to a solution of the harmonic Einstein equation -- the condition is then preserved by the evolution equations as we discuss later.
For the spatial vierbein components we evolve as before using,
\be
0 = {F}_{{e}_{ij}} \equiv \partial_t e_{ij} - P_{ij} - \partial_i V_j 
\ee
and for the new variables we evolve using the same equation but adding this consistency condition as,
\be
0 = {F}_{\tilde{e}_{ij}} \equiv \partial_t \tilde{e}_{ij} - P_{ij} -  Q_{ijt} - \partial_j V_i 
\ee
where now $\{ij\} \in \{ xy, yz, zx \}$.
Finally for the evolution of the time-space vierbein components, $V_i$, we modify the evolution equation given by the momenta, $\partial_t V_j = P_{j} + \partial_j \phi$, by adding the vector constraint as,
\be
0 = {F}_{{V}_{x}} \equiv  \partial_t V_x - P_{x} - \partial_{x} \phi - \frac{\lambda^2}{1 - \lambda^2} \left( \left. \xi_x \right|_{e_{xy} \to \tilde{e}_{xy} \; , \; e_{yz} \to \tilde{e}_{yz} }\right)
\ee
and we obtain expressions for $ {F}_{{V}_{y,z}}$ by cyclicly permuting the indices $x,y$ and $z$.
We take $\lambda > 0$ to be a positive real constant, not equal to one, and in the vector constraint expression above all instances of the variables $e_{xy}$ and $e_{yz}$ are replaced by their auxiliary counterparts. We note that in the expression above $e_{zx}$ is not replaced, although it will be in $ {F}_{{V}_{y,z}}$.

Again we emphasize that while adding the vector constraint to the first order system changes it, in the harmonic formulation we should ensure the vector constraint and its time derivative vanishes in the initial data, and provided the scalar constraint also vanishes, then it  remains zero under evolution. Thus while the addition of this constraint will change the general solution to the first order system, it will not affect a solution that initially satisfies $\xi$ vanishing, and hence a solution to dRGT massive gravity.

This procedure works to give a strongly hyperbolic system on the Minkowski background as we detail shortly. However it is unclear to us that it is the simplest such formulation. One might also wonder whether one needed to pass to the harmonic formulation of the theory, or whether with similar modifications the original formulation could be rendered well-posed.
These are interesting questions we leave for future work. We now turn to the task of demonstrating this system is indeed well-posed, by which we mean strongly hyperbolic.

In order to proceed let us be very explicit regarding the arrangement of the variables in the first order system, and the ordering of the equations. Schematically the vector of variables now takes the form,
\be
\matvec{u} = \left(
\phi, P_i, P_{ij}, V_i , e_{ij}, Q_{ij\mu} , \tilde{e}_{ij}
\right) \; .
\ee
Explicitly we order these 34 variables in the vector as follows;
\be
\label{eq:uvec}
\matvec{u} = \Big(
&& \phi, P_x, P_y, P_z, P_{xx},  P_{yy},  P_{zz},  P_{xy}, P_{yz},  P_{zx}, \nl
&&    V_x, V_y, V_z , e_{xx},e_{yy},e_{zz},e_{xy},e_{yz},e_{zx}, \nl
&&   Q_{xyy} ,Q_{xyx} ,Q_{yzz} ,Q_{yzy} ,Q_{zxx} ,Q_{zxz} ,\nl
&&   Q_{xyt} ,Q_{yzt} ,Q_{zxt} ,\nl
&&   Q_{xyz} ,Q_{yzx} ,Q_{zxy} ,\nl
&&   \tilde{e}_{xy}, \tilde{e}_{yz}, \tilde{e}_{zx}
\Big) \; .
\ee
We choose to order the 34 evolutions equations as the vector of equations,
\be
\matvec{F}[\matvec{u}] = \Big(
&& \partial_t \mathcal{S} , \mathcal{E}^H_{tx} , \mathcal{E}^H_{ty} , \mathcal{E}^H_{tz} 
, \mathcal{E}^H_{xx} , \mathcal{E}^H_{yy} , \mathcal{E}^H_{zz} , \mathcal{E}^H_{xy} , \mathcal{E}^H_{yz} , \mathcal{E}^H_{zx} \nl
&&   {F}_{{V}_{x}} , {F}_{{V}_{y}} , {F}_{{V}_{z}} , 
{F}_{{e}_{xx}}, {F}_{{e}_{yy}},{F}_{{e}_{zz}},
{F}_{{e}_{xy}},{F}_{{e}_{yz}},{F}_{{e}_{zx}},
\nl
&&  
F_{Q_{xyy}}, F_{Q_{xyx}}, F_{Q_{yzz}}, F_{Q_{yzy}}, F_{Q_{zxx}}, F_{Q_{zxz}} ,\nl
&& 
\tilde{F}_x, \tilde{F}_y, \tilde{F}_z, \tilde{F}'_x, \tilde{F}'_y, \tilde{F}'_z,
\nl
&& {F}_{\tilde{e}_{xy}},{F}_{\tilde{e}_{yz}},{F}_{\tilde{e}_{zx}}
\Big) \; .
\ee
From these we construct the first order system as,
\be
\label{eq:firstordersystem}
\matvec{F}[\matvec{u}] = \mat{A}[\matvec{u}]  \cdot \partial_t \matvec{u} + \matvec{P}^i[\matvec{u}]  \partial_i u + \matvec{C}[\matvec{u}] = 0
\ee
using the fact that, as noted above, the system is linear in derivatives. From this we construct the matrix, $\mat{M}[\hat{k}_i] = -\matvec{A}^{-1} \matvec{P}^i \hat{k}_i$ that characterizes its well-posedness, where again $\hat{k}_i$ is a unit wavevector. 
If the system is strongly hyperbolic about a background \matvec{u} then $\mat{M}$ should everywhere have real eigenvalues that smoothly depend on the spacetime point and a complete set of eigenvectors for any $\hat{k}_i$. However since we have preserved rotational invariance in our choice of variables and equations, we may assess hyperbolicity focussing only on $\mat{M}_x \equiv \mat{M}[(1,0,0)]$ without loss of generality provided we consider the set of backgrounds related by rotations.

The vector of data $\matvec{u}$ determines the background about which we ask whether the system is well-posed. If this background data is to represent a solution to the harmonic dRGT system, it should solve the initial data consistency conditions $I_i = J = 0$ given in equations~\eqref{eq:consistency1} and~\eqref{eq:consistency2}, and that in equation~\eqref{eq:consistencyQijt} which together are spatial differential conditions on $\matvec{u}$, and further should have that the auxiliary spatial off-diagonal vierbein variables are equal to those of the vierbein, so $\tilde{e}_{xy} = e_{xy}$ and similarly for $\tilde{e}_{yz}$ and $\tilde{e}_{zx}$. Furthermore it should satisfy the scalar constraint equation, which we may write as $\mathcal{S}[ \matvec{u} ]$ and is purely algebraic in the data $\matvec{u}$.
We note that in order to be a solution to the dRGT system itself, we would also require that $\xi_\mu = 0$ together with the Hamiltonian and momentum constraints $\dot{\xi}_\mu = 0$.
A full discussion of the constraints that must be placed on the initial data of this first order system in order to have a solution of the harmonic dRGT equations is given later in Section~\ref{sec:initialdata}, and further it is shown that these continue to hold after time evolution. 

Firstly we consider the theory about the Minkowski background, so that the only non-zero elements of the vector $\matvec{u}$ are $\phi = -1$ and $e_{xx} = e_{yy} = e_{zz} = 1$. Then we find that the matrix $\mat{A}$ has determinant,
\be
\det\left(\mat{A}\right) =  \frac{3 m^2}{(1 - \lambda^2)^3}
\ee
and since $\lambda^2 \ne 1$ then it is invertible. Now computing the Jordan form of $\mat{M}_x$ we find the (invertible) matrix $\mat{S}$ such that,
\be
\label{eq:firstorderwellpos}
\mathrm{Jordan}\left[ \mat{M}_x \right] = \mat{S}^{-1} \cdot \mat{M}_x \cdot \mat{S} = \left(
\begin{array}{c|c|c|c|c}
-\mat{1}_9 & 0 & 0 & 0 & 0 \\ \hline
& + \mat{1}_9 & 0 & 0 &0  \\ \hline
& & + \mat{0}_{10} & 0 & 0 \\  \hline
& & & 
\begin{array}{cccc}
- {\lambda} & 0 & 0 & 0 \\
& - {\lambda} & 0 & 0 \\
& & + {\lambda}  & 0 \\
& & & + {\lambda}  \\
\end{array}
&0
 \\  \hline
& & & &
\begin{array}{cc}
- {\mu} & 0  \\
& + {\mu} 
\end{array}
\end{array}
\right)
\ee
so that indeed (for real $\mu, \lambda > 0$) the eigenvalues are real, and the matrix has been diagonalized, and hence has a complete set of eigenvectors given by the columns of the matrix $\mat{S}$ with eigenvalues given by the values on the diagonal above. For completeness we give both the matrix $\mat{M}_x$ and the matrix $\mat{S}$ in the Appendix~\ref{app:details}. 

One can confirm that if we take a general $\mat{M}[{k}_i]$, rather than just that associated to a unit wavevector in the $x$-direction, we obtain the same result but now the eigenvalues are all multiplied by $\sqrt{ k_i k_j \delta^{ij} }$, which reflects that the rotational symmetry of the underlying theory on the Minkowski background is preserved in our first order formulation.

The eigenvalues are what we would expect for the harmonic formulation. The nine values $\pm 1$ comprise the five physical graviton modes, together with the four constraint violating modes where $\xi_\mu \ne 0$. The ten zero eigenmodes are associated to the extension of the system when putting it in first order form, namely introducing extra variables for spatial derivatives -- just as in the wave equation example in section~\ref{sec:example}. In addition we have the eigenvalues $\pm {\lambda}$ and $\pm {\mu}$ associated both to these extra variables and the additional auxiliary off-diagonal vierbein components $\tilde{e}_{xy}$,$\tilde{e}_{yz}$,$\tilde{e}_{zx}$ that we have needed to add to obtain a well-posed system. The fact that these depend on the constants $\lambda$ and $\mu$ that we have introduced helps to identify that they are not the physical eigenvalues. 

Having seen that this formulation is well-posed on the Minkowski background, we now wish to understand whether it remains well-posed when we deform away from Minkowski spacetime. 
Firstly the matrix $\mat{A}$ should be invertible, otherwise one may not solve the first order system~\eqref{eq:firstordersystem} to evolve the variables. We have seen that for Minkowski this is invertible.
By continuity it should remain so in a sufficiently small neighbourhood of Minkowski.
More generally $\mat{A}$ should generically be invertible, although there may potentially be codimension one loci in the phase space where its determinant vanishes, and it fails to be.
While the  actual expression for the determinant of $\mat{A}$ is extremely complicated and we have not been able to properly analyse it,
from the example of the wave equation earlier, we might expect invertibility of $\mat{A}$  failing to be associated with constant $t$ hypersurfaces becoming null. Ideally we would give a completely general analysis of the eigenvalues and vectors of the matrix $\mat{M}_x$ on a general configuration. While we have not been able to do this to date, and leave such a complete treatment for future analysis, we are able to argue that for a background in a generic neighbourhood of Minkowski spacetime 
and at a generic spacetime point and wavevector direction $\hat{k}_i$ then $\mat{M}[\hat{k}_i]$ does indeed have a complete set of eigenvectors with real eigenvalues.
In the remainder of the paper we will give the details of this argument. However as we will discuss later, at a given point there will generically exist a set of measure zero of special directions $\hat{k}_i$ where we are not currently able to prove completeness of the eigenvectors -- although have no evidence to the contrary either. Hence we cannot provide a full proof of well-posedness. 

Before proceeding it is worth noting what the challenge is to showing completeness of the eigenvectors. 
Suppose we have a real matrix $\mat{X}$ that has distinct (i.e. non-degenerate) real eigenvalues and a diagonal Jordan form -- meaning it has a complete set of eigenvectors.
Consider a deformation $\mat{X}' = \mat{X} + \epsilon \mat{Y}$ to this given by the real matrix $\mat{Y}$ and real deformation parameter $\epsilon$. Then $\mat{X}'$ shares these same features of real eigenvalues and a complete set of eigenvectors for some neighbourhood of zero in the deformation $\epsilon$.
Let us understand why.

Since eigenvalues are solutions of the characteristic polynomial, which here has real coefficients (since the matrices $\mat{X}$ and $\mat{Y}$ are real), then any complex eigenvalues must come in conjugate pairs. A deformation that results in complex eigenvalues then must be such that (at least) two real eigenvalues are deformed to coincide, and then continuing the deformation become a complex conjugate pair. 
Thus we see that given the real matrix $\mat{X}$ has distinct eigenvalues, then for $\epsilon$ sufficiently close to zero, these must remain distinct, and hence real.

However we are interested in $\mat{M}_x$ having more than just real eigenvalues; it must also have a complete set of eigenvectors. For a matrix $\mat{X}$ not to have a complete set of eigenvectors, its Jordan form must have a two-by-two or larger block with entries off the diagonal. However the diagonal entries in any Jordon block must have the same value. Since these diagonal entries correspond to the eigenvalues, we see that failure to have a complete set of eigenvectors requires degenerate \emph{defective} eigenvalues, and therefore degenerate roots of the characteristic polynomial.\footnote{
Recall that a square matrix that lacks a complete set of eigenvectors is termed defective. The diagonal values of a non-diagonal Jordan block for such a matrix are called defective eigenvalues. They correspond to eigenvectors, but with lesser multiplicity than the number of these diagonal values.}
Hence again as the eigenvalues must remain distinct for sufficiently small $\epsilon$, it must be that the diagonal Jordan form is preserved, and hence the deformed matrix retains a complete set of eigenvalues.

Returning to our first order system we now see that the challenge in understanding whether it remains well-posed as we deform away from the Minkowski background is the challenge of understanding the blocks associated to degenerate eigenvalues. If the perturbation of $\mat{M}_x$ introduces any off diagonal term into the upper triangular part of these blocks, then that deformed Jordan form will imply a lack of eigenvectors -- if that were the case then some eigenvectors on the Minkowski background would fail to remain eigenvectors when the system is deformed. A toy model for this situation is the matrix, 
$\left(
\begin{array}{cc}
1 & \epsilon \\
0 & 1
\end{array}
\right)$
which for $\epsilon = 0$ has the two eigenvectors $(1,0)$ and $(0,1)$, but for any non-zero $\epsilon$ only has the first of these.

As we will discuss shortly, for a general background the 9-fold degeneracy of the wavemodes of the harmonic Einstein system is broken to a 6-fold degeneracy. We will term these the degenerate `wavemode' eigenvalues and eigenvectors. The 10 zero eigenvectors remain, and the degeneracy of the block involving $\lambda$ is broken if the auxiliary variables $\tilde{e}_{xy}$, $\tilde{e}_{yz}$, $\tilde{e}_{zx}$ are not equal to their counterpart physical vierbein components ${e}_{xy}$, ${e}_{yz}$, ${e}_{zx}$ but unbroken otherwise. This generic breaking can be straightforwardly seen analytically in special cases of backgrounds, and we discuss such an example background shortly.\footnote{It may also be confirmed by evaluating the Jordan form of $\mat{M}_x$ where the data of the background $\matvec{u}$ are simply given random numerical values (close to their Minkowski values).}
However before we go on to discuss this eigenvalue splitting we will first show that the zero eigenvalues and vectors remain for a general background, as do those eigenvectors with eigenvalue $\pm \mu$, and further that there are always 6 degenerate eigenvalues and eigenvectors governed by the inverse metric.
Then we consider a simple background which gives a generic splitting. Then following this we will go on to analyse the well-posedness near Minkowski spacetime. 
\\

\section{Zero eigenvalues for generic backgrounds}
\label{sec:zeromodes}

We now show that the ten eigenvectors with zero eigenvalue that we have seen on the Minkowski background, and are associated to extending to the first order system by introducing derivative variables, remain on a general background. 
Assuming we are on a generic background so that $\mat{A}$ is invertible, then zero eigenvectors of $\mat{M}[ \hat{k}_i ] = -\mat{A}^{-1} \cdot \mat{P}^i \hat{k}_i$ will be in the kernel of $ \mat{P}^i \hat{k}_i$. Since the theory, and our formulation, have a covariance under spatial rotations, then w.l.o.g. we may consider the case that $\hat{k}_i = (1,0,0)$, and thus consider the kernel of $\mat{P}^x$. 
For a background $\matvec{u}$ at some spacetime point the other wavevector directions can be studied by considering a new background generated by rotating $\matvec{u}$ about that point.

Let us examine  the equations $\matvec{E}[\matvec{u}]$ whose $x$-derivative terms give rise to $\mat{P}^x$. One can see from the discussion above that many of these equations are independent of the background; in particular this comprises \emph{most} of the evolution equations for the derivative variables that have been added. Specifically the equations that do depend on the background are the 9  harmonic Einstein equations $\mathcal{E}^H_{ti}$ and $\mathcal{E}^H_{ij}$ as one would expect, together with the 3 evolution equations for $V_i$. This is less obvious, as naively these simply follow from the definition of the momentum variables, but since we have modified these equations by the term $\frac{\lambda^2}{1-\lambda^2} \xi_i$, and the vector constraint depends on the background, these three evolution equations do. All the remaining $34 - 9 - 3 = 22$ equations have no background dependence, and  all are evolution equations for the derivative variables, except for $\partial_t \mathcal{S}$. While this is a complicated equation, since $\mathcal{S}$ is purely algebraic in the variables $\matvec{u}$, then $\partial_t \mathcal{S}$ only involves time derivatives and hence doesn't contribute to $\mat{P}^x$.

Thus in short, $\mat{P}^x$ has 22 rows that don't depend on the background. We may delete the 12 rows that do depend on the background, keeping those that don't; we keep the first row from $\partial_t \mathcal{S}$, and take the 21 rows from row 14 onwards, and  arrive at the 22 by 34 matrix which we may denote $\mat{\tilde{P}}_x$ -- this is given explicitly in Appendix \ref{app:details}. The rank may be simply computed and we find $\mathrm{Rank}\left[ \mat{\tilde{P}}_x \right] = 12$, showing that $22-12 = 10$ of the rows of $\mat{\tilde{P}}_x$ are linear combinations of the others. This implies that the kernel of the full matrix $\mat{{P}}_x$ is at least 10 dimensional.\footnote{
Since the rank of $\mat{\tilde{P}}_x$ is 12, adding back the 12 background dependent rows to obtain the full 34 by 34 matrix $\mat{P}^x$, we see the rank of $\mat{P}^x$ can be at most $\mathrm{Rank}\left[ \mat{{P}}_x \right] \le 24$. Since the dimension of the vector space is 34, the rank-nullity theorem implies,
\be
\mathrm{Rank}\left[ \mat{{P}}_x \right] + \mathrm{Ker}\left[ \mat{{P}}_x \right] = 34
\ee
and hence we learn the kernel of the full matrix $\mat{{P}}_x$ must be at least $10$.
}
Thus we see that we should have at least 10 eigenvectors with zero eigenvalue of $\mat{M}_x$. Since on a Minkowski background we find precisely 10, this implies that generically we have 10 zero eigenvectors.

\section{Non-zero eigenvalues for generic backgrounds}
\label{sec:nonzeromodes}

For  the non-zero eigenvectors we may solve a number of the linear conditions in the eigenvector equation and reduce it to a 14 by 14 system. Again w.l.o.g. we may consider the case that $\hat{k}_i = (1,0,0)$, and thus consider the eigenvectors of $\mat{M}_x$. We consider the eigenvector condition, $\matvec{Q} \equiv \mat{M}_x \cdot \matvec{v} - \Lambda \matvec{v}$.  Writing the eigenvector as,
\be
\label{eq:eigenvecansatz}
\matvec{v} = \Big(
&& \delta\phi, \delta P_x, \delta P_y, \delta P_z, \delta P_{xx},  \delta P_{yy},  \delta P_{zz},  \delta P_{xy}, \delta P_{yz},   \delta P_{zx},    \delta V_x, \delta V_y, \delta V_z , \delta e_{xx}, \delta e_{yy}, \delta e_{zz},\delta e_{xy},\delta e_{yz},\delta e_{zx}, \nl
&&   \delta Q_{xyy} ,\delta Q_{xyx} ,\delta Q_{yzz} ,\delta Q_{yzy} ,\delta Q_{zxx} ,\delta Q_{zxz} , \delta Q_{xyt} ,\delta Q_{yzt} ,\delta Q_{zxt} ,  \delta Q_{xyz} ,\delta Q_{yzx} ,\delta Q_{zxy} ,   \delta \tilde{e}_{xy}, \delta  \tilde{e}_{yz}, \delta \tilde{e}_{zx}
\Big)
\ee
we may solve a number of linear conditions by taking,
\be
\label{eq:ansatznonzero}
  \delta e_{xx} &=&  + \frac{\delta V_x}{\Lambda} \; , \quad  \delta e_{xy} =   - \frac{\delta V_y}{\Lambda} \; , \quad  \delta \tilde{e}_{zx} =  +  \frac{\delta V_z}{\Lambda} \nl
\delta Q_{xyy} &=& \frac{\delta P_{yy}}{\Lambda} 
\; , \quad  
\delta Q_{xyx} =  \frac{\delta P_{y}}{\Lambda^2} +  \frac{\delta P_{xy}}{\Lambda}
\; , \quad 
\delta Q_{zxx} =   -\frac{\delta P_{zx}}{\Lambda} 
\; , \quad
\delta Q_{zxz} = -\frac{\delta P_{zz}}{\Lambda} 
\nl
\delta Q_{xyt} &=& \frac{\delta P_{y}}{\Lambda} 
\; , \quad
\delta Q_{zxt} =  -\frac{\delta P_{z}}{\Lambda} 
\; , \quad
 \delta Q_{zxy} =  -\frac{\delta P_{yz}}{\Lambda}  \nl
 \delta Q_{xyz} &=& + \frac{\left( \delta P_{yz} +  \delta Q_{yzt} \right)}{\Lambda} \; ,\quad \delta Q_{yzx} = + \frac{ \delta Q_{yzt}}{\Lambda} \nl
 0 & = & \delta e_{yy} = \delta e_{zz} = \delta e_{yz} = \delta e_{zx} = 
 \delta Q_{yzz} = \delta Q_{yzy}  =\delta Q_{zxz}  = \delta Q_{xyt}   \; .
\ee
Having solved these linear relations  we may reduce the system to one in terms of the 14-dimensional vector,
\be
\hat{\matvec{v}} = \Big(
&& \delta\phi, \delta P_x, \delta P_y, \delta P_z, \delta P_{xx},  \delta P_{yy},  \delta P_{zz},  \delta P_{xy}, \delta P_{yz},   \delta P_{zx},    \delta V_x, \delta V_y, \delta V_z ,\delta Q_{yzt} 
\Big)
\ee
and the 14 non-trivial remaining equations $\matvec{\hat{Q}}$, and then re-cast it as the smaller system,
\be
\hat{\matvec{Q}} \equiv \hat{\mat{M}}(\Lambda) \cdot \hat{\matvec{v}} = 0
\ee
where $\hat{\mat{M}}$ is a 14 by 14 matrix that now depends non-linearly on $\Lambda$. 
Here,
\be
\matvec{Q} = \left( \hat{\matvec{Q}}_{1,\ldots 13}, 0, \ldots , 0, \hat{\matvec{Q}}_{14} , 0, \ldots, 0 \right) 
\ee
so that the first 13 components of $\hat{\matvec{Q}}$ come from the first components of $\matvec{Q}$ and the last comes from the 27th component, which is associated to the evolution of the variable $Q_{yzt}$.

In order to have non-trivial eigenvectors we therefore require that the eigenvalue $\Lambda$ solves $\det( \hat{\mat{M}} ) = 0$. Three of the linear conditions take a simple form; these are,
\be
\label{eq:lincondition}
0 & = & \delta Q_{yzt} \left( \Lambda^2 - \mu^2 \right) \nl
0& = & \frac{\delta V_y}{\Lambda} \left. \left( \frac{\lambda^2}{1 - \lambda^2} (E^{-1})^{\mu\nu} \hat{k}_\mu \hat{k}_\nu - \Lambda^2 \right) \right|_{e_{yz} \to \tilde{e}_{yz}, e_{zx} \to \tilde{e}_{zx}} \nl
0 & =& \frac{\delta V_z}{\Lambda} \left. \left( \frac{\lambda^2}{1 - \lambda^2} (E^{-1})^{\mu\nu} \hat{k}_\mu \hat{k}_\nu - \Lambda^2 \right) \right|_{e_{xy} \to \tilde{e}_{xy}, e_{zx} \to \tilde{e}_{zx}} 
\ee
and these constrain $\delta Q_{yzt}$ to vanish if we do not select $\Lambda = \pm \mu$, and likewise they require $\delta V_y$ and $\delta V_z$ to vanish unless $\Lambda$ is chosen appropriately so the relevant factors above vanish.

We now examine some of the non-zero eigenvalues and their eigenvectors for a generic background. In particular we will show that as for the zero-modes discussed above, on any background there are eigenvectors with eigenvalues $\pm \mu$. 
We then show one of the main results of this paper, namely that there are always at least 6 degenerate modes that are simply governed by the inverse metric. Later we will see that these physically correspond to the spin-two graviton mode, as well as the four constraint violating modes of the harmonic formulation. 
Finally we show that in the special case of interest, $\tilde{e}_{ij} = e_{ij}$, on a generic background there are two pairs of eigenvectors -- the `$\lambda$' eigenvectors -- with degenerate real eigenvalues that depend on $\lambda$, just as for the Minkowski background. 
With the accumulated information about these non-zero eigenvalues and eigenvectors and their degeneracies on a general background, we consider an analytically tractable background where we can explicitly solve for all eigenvectors and eigenvalues, and which gives precisely the degeneracies we have seen generically occur. 
We then go on to examine well-posedness in a neighbourhood of Minkowski in the following section~\ref{sec:wellposedMink}.

\subsection{$\mu$-eigenvalues}
\label{sec:mumodes}

Having chosen $\mu > 0$ with $\mu \ne 1$,  then the two eigenvectors on the Minkowski background with eigenvalues $\pm \mu$ are distinct from all the others. 
Interestingly on a generic background the eigenvalues  remain as $\pm \mu$ as we now show. As we noted above, it is only for this special value that $ \delta Q_{yzt}$ may be non-zero.

In our general ansatz above we further take,
\be
\delta V_x &=& - \frac{\delta \phi}{\Lambda} \; , \quad \delta V_y = \delta V_z = 0
\ee
and then choosing $\Lambda = \pm \mu$ we solve the linear conditions in equation~\eqref{eq:lincondition} and we may rewrite the system in terms of a 10 by 10 matrix $\mat{W}$ acting on the 10-dimensional vector, $\vec{u}$, with the system being sourced by $ \delta Q_{yzt}$ as,
\be
 \mat{W} \cdot \matvec{{u}} = \matvec{S} \, \delta Q_{yzt} \; , \quad \matvec{u} = \left(  \delta\phi, \delta P_x, \delta P_y, \delta P_z, \delta P_{xx},  \delta P_{yy},  \delta P_{zz},  \delta P_{xy}, \delta P_{yz},   \delta P_{zx} \right) 
\ee
where $\matvec{S}$ is a 10 component vector, and like $\mat{W}$, it  depends on the  background.
One may simply compute the determinant of $\mat{W}$ on the Minkowski background, finding it is non-zero (providing $\mu > 0$ and is not equal to one) with value,
\be
\det\left( \mat{W} \right) = -\frac{3 (1 - \mu^2)^9}{\mu^8} \; .
\ee
Hence in a neighbourhood of Minkowski this determinant will remain non-zero, and indeed on any generic background it will be non-zero -- it may vanish on some lower co-dimension subspace of backgrounds, but generically will not vanish. Thus for a generic background we will be able to invert the linear system to solve for $\matvec{u}$ in terms of the free data $\delta Q_{yzt}$. Thus we see that for both eigenvalues $\Lambda = \pm \mu$ an eigenvector exists for generic backgrounds.

\subsection{Degenerate wavemode eigenvectors}
\label{sec:physmodes}

We now demonstrate that there is at least a 6-fold degeneracy associated to the wave propagation of the harmonic Einstein equations, and that the eigenvalues are computed from the metric lightcone. More precisely we claim there are at least 12 eigenvectors, $\matvec{v}_{(\pm A)}$ with $A = 1,\ldots, 6$ such that,
\be
\label{eq:wavemode1}
\mat{M}_x \cdot \matvec{v}_{(\pm A)} =  \Lambda_{\pm} \matvec{v}_{(\pm A)}
\ee
where the eigenvalue is determined from the quadratic condition,
\be
\label{eq:wavemode2}
\hat{k}_\mu = \left( + \Lambda, \hat{k}_i \right) = \left( + \Lambda, 1, 0, 0 \right) \; , \qquad g^{\mu\nu} \hat{k}_\mu \hat{k}_\nu = 0
\ee
with $\Lambda_{\pm}$ being the two roots, which assuming $g_{\mu\nu}$ is Lorentzian, must be real. Thus these modes propagate on the usual metric lightcone, as for the graviton wavemodes in conventional GR. We will term these the degenerate `wavemode' eigenvectors.

Now generically for these modes we will have $\Lambda^2 \ne \mu^2$ and hence taking our ansatz above consistency requires $\delta Q_{yzt} = 0$. For $\delta V_i$ we take a similar form to that for the $\mu$-modes above. Hence we have,
\be
\delta V_x &=& - \frac{\delta \phi}{\Lambda} \; , \quad \delta V_y = \delta V_z = 0 \; , \quad \delta Q_{yzt} = 0
\ee
which solves the linear conditions in equation~\eqref{eq:lincondition},
and we may write our system again as a 10-dimensional one with,
\be
\label{eq:wavemodeansatz}
\hat{\matvec{Q}}' \equiv \mat{W}' \cdot \matvec{\hat{v}} = 0 \; , \quad \matvec{u} = \left(  \delta\phi, \delta P_x, \delta P_y, \delta P_z, \delta P_{xx},  \delta P_{yy},  \delta P_{zz},  \delta P_{xy}, \delta P_{yz},   \delta P_{zx} \right) 
\ee
where $\hat{\matvec{Q}}'$ are the non-trivial equations given by the first 10 components of $\matvec{Q}$, and $\mat{W}'$ is a 10 by 10 matrix depending on the background and non-linearly on $\Lambda$. 
The key point is that if $\det \mat{W}' \neq 0$ then the only solution to the above is $\matvec{\hat{v}} = {\mat{W}'}^{-1} \cdot \matvec{0} = \matvec{0}$. Our aim is now to show that for the particular choice \eqref{eq:wavemode2}, the matrix $\mat{W}'$ is not full rank, and in fact that its kernel is at least six dimensional.

First, one can directly check that with this choice of the eigenvalue and ansatz,
\be 
\label{eq:condvanish1}
\hat{Q}'_6 = \hat{Q}'_7 = \hat{Q}'_9 = 0
\ee
so that three rows of the $\mat{W}'$ matrix vanish outright. These come from the equations for $\delta P_{ij}$ variables that do not involve $x$-derivatives, namely $\hat{\mathcal{E}}^H_{yy}, \hat{\mathcal{E}}^H_{zz}$ and $\hat{\mathcal{E}}^H_{yz}$, where $\hat{\mathcal{E}}^H_{\mu\nu}$ are the terms in the harmonic Einstein condition given earlier in equation~\eqref{eq:harmK} that depend on derivatives of our variables, so,
\be
\hat{\mathcal{E}}^H_{\mu\nu} &=& \mathcal{A}^{\sigma\alpha\rho\beta}_{\mu\nu} \partial_\sigma K_{\alpha\rho\beta} + \mathcal{C}^{\sigma\rho\delta(\alpha\beta)\gamma}_{\mu\nu} K_{\sigma\rho\delta} \partial_{(\alpha} E_{\beta)\gamma}  \; .
\ee
Next, we find the linear relations
\ba
\label{eq:condvanish2}
0 &=& \hat{Q}'_4-\Lambda\hat{Q}'_{10} \nl
0 &=& \hat{Q}'_3-\Lambda\hat{Q}'_{8} 
\ea
which implies that the four rows $W'_{3,i},W'_{4,i},W'_{8,i},W'_{10,i}$ span only a two-dimensional space. In terms of the underlying equations, these correspond to the derivative terms occurring in the linear combinations $\hat{\mathcal{E}}^H_{tz} - \Lambda \hat{\mathcal{E}}^H_{zx}$ and $\hat{\mathcal{E}}^H_{ty} - \Lambda \hat{\mathcal{E}}^H_{xy}$. 

We need one more linear relation on the system of equations. This, somewhat miraculously, comes from the fact that using the relations above,
\ba
\label{eq:condvanish3}
&2&\Lambda\left(\Lambda (g^{-1})^{tt}+(g^{-1})^{tx}\right)\hat{Q}'_2 + 2\left(\Lambda (g^{-1})^{ty}+(g^{-1})^{xy}\right)\hat{Q}'_3 + \nl +&2&\left(\Lambda (g^{-1})^{tz}+(g^{-1})^{zx}\right)\hat{Q}'_4  - \Lambda\left(\Lambda^2 (g^{-1})^{tt} - (g^{-1})^{xx}\right)\hat{Q}'_5 = 0
\ea
which originates from the derivative terms in the linear combination of equations $\hat{\mathcal{E}}^H_{\mu\nu}\hat{k}^\mu\hat{k}^\nu$ for this ansatz.
This  can checked explicitly, and we refer the reader to the accompanying Mathematica notebook for the calculation~\cite{Notebook}. Note that while we have not directly imposed  $\hat{\mathcal{E}}^H_{tt}$, it can be determined from the other Einstein equations through the scalar constraint  $g^{\mu\nu} \hat{\mathcal{E}}^H_{\mu\nu}\sim 0$ up to terms that don't involve derivatives. This then determines $\hat{Q}'_5$ in terms of the equations $\hat{Q}'_2,\hat{Q}'_3,\hat{Q}'_4$. Thus we see that the equations $\hat{Q}'_{5, \ldots, 10}$ vanish or are linearly related to the first four equations, $\hat{Q}'_{1,2,3,4}$. Hence for these eigenvalues we see that solving these four equations implies the full eigenvector system is solved. 

The six conditions \eqref{eq:condvanish1}, \eqref{eq:condvanish2} and \eqref{eq:condvanish3} taken together now imply that the kernel of the matrix $\mat{W}'$ is at least six dimensional, or in other words that there are at least six linearly independent solutions to the equation \eqref{eq:wavemode1} for $\Lambda_+$ and $\Lambda_-$. As we shall see shortly for a particular analytic background, generically there are \emph{at most} six such eigenvectors.  Of course, on particular backgrounds there may be more than six solutions: in Minkowski space there are nine.

\subsection{Degenerate $\lambda$ eigenvectors for $\tilde{e}_{ij} = e_{ij}$}
\label{sec:lambdamodes}

There are four further modes that depend explicitly on $\lambda$ which we may understand on a generic background, and further which form two degenerate pairs when we take exactly $\tilde{e}_{ij} = e_{ij}$, which is the class of solutions we are actually interested in. These are governed only by the vierbein of the background (and not derivatives) but in a more complicated manner than that of the 6 degenerate wavemodes modes. 

Here we generically have $\Lambda^2 \ne \mu^2$ and hence again we take our general ansatz for non-zero modes, and we must set $\delta Q_{yzt} = 0$. 
Now consider the linear conditions on $\delta V_y$ and $\delta V_z$ in equation~\eqref{eq:lincondition}.
The factors $\left(\frac{\lambda^2}{1 - \lambda^2} (E^{-1})^{\mu\nu} \hat{k}_\mu \hat{k}_\nu - \Lambda^2\right)$ in these expressions are quadratic in $\Lambda$, and generically are different assuming that $\tilde{e}_{ij} \ne e_{ij}$. In this case we may choose the eigenvalue $\Lambda$ appropriately so that one or other of these factors vanishes, and hence either $\delta V_y$ or $\delta V_z$ may be non-zero, while the other must be set to zero. 
This will then yield 4 linearly independent eigenvectors; two with $\delta V_y = 0$ and $\Lambda$ determined from the two roots of the quadratic with the replacements $e_{xy} \to \tilde{e}_{xy}$ and $e_{zx} \to \tilde{e}_{zx}$, and the other two with $\delta V_z = 0$ and $\Lambda$ given as the two roots of the same expression but now with the replacements $e_{yz} \to \tilde{e}_{yz}$ and $e_{zx} \to \tilde{e}_{zx}$. 

On the other hand if we have $\tilde{e}_{ij} = e_{ij}$ then choosing $\Lambda$ to solve the quadratic $\left(\frac{\lambda^2}{1 - \lambda^2} (E^{-1})^{\mu\nu} \hat{k}_\mu \hat{k}_\nu - \Lambda^2\right)$ then the linear conditions in equation~\eqref{eq:lincondition} are trivially solved and both $\delta V_y$ and $\delta V_z$ may be non-zero. This will yield two degenerate pairs of eigenvalues (one for each root of the quadratic), with each linearly independent pair parameterized by $\delta V_y$ and $\delta V_z$.

We should check that in both these cases the corresponding eigenvectors exist. The linear system reduces to,
\be
 \mat{W}'' \cdot \matvec{u}' = \mat{U} \cdot \left( \begin{array}{c} \delta V_y \\ \delta V_z \end{array} \right) 
 \; , \quad \matvec{u}' = \left(  \delta\phi, \delta P_x, \delta P_y, \delta P_z, \delta P_{xx},  \delta P_{yy},  \delta P_{zz},  \delta P_{xy}, \delta P_{yz},   \delta P_{zx}, \delta V_x \right) 
\ee 
so that $\mat{W}''$ is an 11 by 11 matrix, and $\mat{U}$ is a 11 by 2 matrix, both of which depend on $\Lambda$ and the background. 
We find that for the Minkowski background,
  \be
  \det\left(  \mat{W}'' \right) =\frac{(1-\Lambda^2)^9}{\Lambda^7}
  \ee
and these eigenvalues are $\Lambda = \pm \lambda$ as we have seen previously. Hence this determinant is non-zero, assuming as we have that $\lambda \ne 1$, and hence we may solve for $\matvec{u}$, and hence the eigenvector, in terms of the undetermined $\delta V_y$ and/or $\delta V_z$ (depending on whether $\tilde{e}_{ij}$ is equal to $e_{ij}$ or not).
Thus by continuity there exists an open neighbourhood about Minkowski where these eigenvectors exist. Indeed for generic background configurations this determinant will also be non-zero too, although there may be special non-generic points in the space of backgrounds where it does vanish.

\section{An analytic example}
\label{sec:analyticexample}

Using a simple analytic background, we now detail how at a point we obtain a generic pattern of degeneracy breaking for the non-zero eigenvalues, and well-posedness there.
In particular we will see the degeneracies that we have derived for a general background above are the only ones that occur. We will find 6 and only this number of degenerate wavemode eigenvalues governed by the inverse metric, which we will argue includes the spin-two graviton. We will also see that for $\tilde{e}_{ij} = e_{ij}$ an additional two degenerate pairs of eigenvalues occur that depend on $\lambda$.

Consider a point $p$, which w.l.o.g. we choose to be at $x^\mu = 0$, and consider the background given by the vierbein,
\be
E_{\mu\nu} =  \left( 
\begin{array}{c|c}
\phi & 0  \\ \hline
0 & \delta_{ij}
\end{array} 
\right) + O( x^\mu )
\ee
so that $V_i = 0$ and $e_{ij} = \delta_{ij}$ at $p$. Then we take all the derivative variables to vanish at $p$ except for $Q_{xyt}$. Finally we take the auxillary off-diagonal spatial vierbein variables to be vanishing at $p$ except for $\tilde{e}_{xy}$.

We note that the vector constraint $\xi_\mu$ and scalar constraint $\mathcal{S}$ at $p$ depend only algebraically on the data $\matvec{u}$ there, and not on any derivatives of $\matvec{u}$. Hence they may be evaluated and one finds that at $p$ they are,
\be
\left. \xi_\mu \right|_p = 0 \; , \quad \left.\mathcal{S} \right|_p = \frac{-1}{2 \phi^2} \left( 6 m^2\phi ( 1 + \phi) + Q_{xyt}^2 ( 1 - \phi)^2 \right) \; .
\ee
Now we recall that our first order system is an extension of the harmonic Einstein equations. If we require that the background can be a solution for the harmonic Einstein equations then at the point $p$ we require $\tilde{e}_{xy} = 0$ (since $e_{xy} = 0$ here) and furthermore the scalar constraint $\mathcal{S}$ must be satisfied there. From above we see that $\mathcal{S} = 0$ may be solved for $\phi$ assuming that $Q_{xyt}^2$ not too large, specifically that $ Q_{xyt}^2 \le 3m^2/4$. 

There are also the various consistency conditions coming from the definitions of the derivative variables, such as $I_i$, $J$ and equation~\eqref{eq:consistencyQijt} above, that should be satisfied for a solution of the harmonic Einstein equations, but these involve derivatives of the variables $\matvec{u}$, rather than just their value at the point $p$, and so will not concern our current discussion which is localized at $p$.
The fact that $\xi_\mu$ vanishes at $p$ implies that in fact this could be a consistent solution to dRGT there, not just the harmonic Einstein equation. One would also have to ensure that $\dot{\xi}_\mu$ vanished too, but this condition involves derivatives of $\matvec{u}$, and thus is not local to $p$, so again doesn't concern us.

Now we may consider well-posedness at the point $p$. We may observe the splitting of the degenerate eigenvalues by looking at derivatives in the $x$-direction. Explicitly the characteristic polynomial of $\mat{M}_x$ is given by,
\be
P(\Lambda) = c \, \Lambda^{10} P_\lambda(\Lambda) P_\mu(\Lambda) P_{wave}(\Lambda)  
\ee
where $\Lambda$ is the eigenvalue, $c$ is a $\Lambda$ independent coefficient, and the various factors controlling the eigenvalues are explicitly,
\be
P_\lambda(\Lambda) & = &  \left( ( 1 -  \tilde{e}_{xy}^2) \Lambda^2 \phi - \lambda^2 \left(  \phi + \Lambda^2 (1 + \phi) ( 1 -  \tilde{e}_{xy}^2) \right) \right)  \left(  \Lambda^2 \phi - \lambda^2 \left( \phi  + \Lambda^2 (1 + \phi)  \right) \right) \nl
P_\mu(\Lambda) & = & \Lambda^2 - \mu^2 
\ee
with the remaining factor being,
\be
\label{eq:polyphys}
P_{wave}(\Lambda) = \left( \Lambda^2 - \phi^2 \right)^6 \left( 2 \Lambda^2 + \phi (1 - \phi) \right) 
\left( 2 (  Q_{xyt}^2 - 3 m^2) \Lambda^4 - \Lambda^2 \phi \left( 3m^2 - 9m^2 \phi + 4 Q_{xyt}^2 \right) + \phi^2 \left( 2 Q_{xyt}^2 + 3m^2 \phi ( 1 - \phi ) \right) \right) \; . \nl
\ee
We see the factor of $\Lambda^{10}$ associated to the 10 zero eigenvalues and their corresponding eigenvectors that we showed generically exist above. Also evident are the two eigenvalues $\Lambda = \pm \mu$ that we discussed in the previous section. Furthermore the 6 degenerate eigenvalues $\Lambda = \pm \phi$ are simply the wavemode eigenvalues seen above for a general background, with $\Lambda$ simply determined by the metric lightcone condition. Thus the 9 degenerate pairs of eigenvectors with eigenvalue $\pm 1$ associated to the graviton and $\xi_\mu$ constraint violating modes of the harmonic theory for the Minkowski background split to give only 6 degenerate pairs.
Likewise the 4 eigenvalues that are roots of $P_\lambda(\Lambda)$ are precisely those discussed above generally. In particular we see that these are generally distinct, but become two degenerate pairs when $\tilde{e}_{xy} = 0$, and hence in this case, $\tilde{e}_{ij} = e_{ij}$.

For $\hat{k}_i = ( \cos{\theta}, \sin{\theta}\cos{\xi}, \sin{\theta} \sin{\xi})$ in a general direction the characteristic polynomial has a similar form.
We note that since the degenerate wavemode eigenvalues depend only on the vierbein, and not the momentum and derivative variables, they are independent of $\hat{k}_i$ if the metric is isotropic as it is here for this analytic background. 
For a generic direction we see the same splitting as for the $x$ direction, but we note that there are special directions where there is extra degeneracy -- for example for $\hat{k}_i = (0,0,1)$ the factor above going as $\left( 2 \Lambda^2 + \phi (1 - \phi) \right)$ becomes doubly degenerate. The non-degenerate wavemodes do depend explicitly on the direction $\hat{k}_i$, as do the $\lambda$ modes since $\tilde{e}_{xy}$ explicitly picks out a direction. In the case $\tilde{e}_{xy} = 0$, so that $\tilde{e}_{ij} = \tilde{e}_{ij}$ then these $\lambda$ modes become degenerate and depend only on the isotropic vierbein and  not  on $\hat{k}_i$.

The eigenvalues that have split and become non-degenerate must have corresponding eigenvectors. The obstruction to well-posedness is that the degenerate eigenvalues may be defective. However for a general direction $\hat{k}_i$ we observe precisely the 6 degenerate wavemodes we have shown exist for a generic background, and further we have shown that the corresponding set of eigenvectors always exist for these degenerate eigenvalues. Likewise if $\tilde{e}_{xy} = 0$ we have two additional degenerate pairs, but this is generic to any background and we have shown above that the corresponding eigenvectors exist.
Thus for a generic direction $\hat{k}_i$ the matrix $\mat{M}[ \hat{k}_i ]$ has a complete set of eigenvectors, and hence provided the eigenvalues are real then the background is well-posed at the point $p$. We note that as we deform from Minkowski spacetime, the eigenvalues that split and remain non-degenerate  must be real. The degenerate ones may potentially become complex as we deform away from Minkowski. The degenerate wavemodes are determined by the inverse metric, so as usual provided this is Lorentzian they will be real. Since $\phi = -1$ for Minkowski and should be $\phi < 0$ generally, then $\lambda$ modes can only become complex if $\lambda^2 > 1$. Thus choosing $\lambda$ to be smaller than one implies they must also be real. Hence this analytic background gives a well-posed dynamics at $p$ provided that the 3 non-degenerate wavemodes remain non-degenerate and real, and indeed sufficiently near to the Minkowski background they do.

An important consideration is that for non-generic $\hat{k}_i$ we may have additional degeneracy amongst the wavemodes, as for $\hat{k}_i = (0,0,1)$, and one might be concerned that these eigenvalues become defective. 
We may check in this particular case that $\mat{M}[  (0,0,1) ]$ still has a full complement of eigenvectors and indeed it does. 

%
%

In the case $\tilde{e}_{xy} =  0$ the wavemode eigenvectors  correspond to the short wavelength wavemodes of the harmonic Einstein equation. We have seen above in the discussion in section~\ref{sec:harmform} that we expect constraint violating modes to be governed by the metric lightcone, and thus a 4 dimensional subspace of these modes should give these constraint violating modes. It is then natural to wonder which  modes the remaining physical  two dimensions correspond to. We use the term physical here as these will be associated to wavemodes of massive gravity as we now outline.
As discussed earlier around equation~\eqref{eq:deltaxi} these eigenvectors have a physical interpretation as short wavelength linear perturbations, and so we may consider the linear variation of $\xi_\mu$ in the short wavelength limit for an eigenvector as $\delta \xi_\mu = \mat{\Xi}_\mu \cdot \matvec{v}$. 
The background does not break $x$-$y$ rotation symmetry and thus we can ask what spin representation a wavemode corresponds to for a  mode propagating in the $z$ direction.
The 6 degenerate wavemodes may be explicitly described using a similar ansatz to that for the wavemode eigenvectors in section~\ref{sec:physmodes}. The ansatz there was taken for $k_i = (1,0,0)$ and here we are interested in propagation in the $z$-direction, so $k_i = (0,0,1)$. Adapting the ansatz appropriately for this, then on this analytic background we may solve for $\delta \phi$ and $\delta P_i$ in the analog of equation~\eqref{eq:wavemodeansatz} explicitly in terms of the 6 dimensional data $\delta P_{ij}$. We then find,
\be
\delta \xi_\mu = \mat{\Xi}_\mu \cdot \matvec{v} = \left(  \delta P_{xx}+\delta P_{yy}+\delta P_{zz} , \pm \frac{1}{\phi} \delta P_{zx}, \pm \frac{1}{\phi} \delta P_{yz}, \pm \frac{1}{\phi} \delta P_{zz} \right) \; .
\ee
Thus we see that for propogation in the $z$ direction, the two physical modes are orthogonal to the constraint violating ones, so have $\delta P_{zx}, \delta P_{yz}, \delta P_{zz}$ vanishing together with $\delta P_{xx}+\delta P_{yy} = 0$, and  so writing $\delta P_{IJ}$ with $I,J \in \{ x, y \}$ the data is then given by this 2 by 2 traceless matrix, and hence is a spin-2 representation of the rotation group about the $z$ axis.

After the two sets of 9 degenerate wavemode eigenvalues about Minkowski spacetime split to two sets of 6 degenerate eigenvalues, which comprise the constraint violating mode and the spin-2 graviton, there remain 6 non-degenerate eigenvalues. These 6 correspond to the continuation of the spin-0 and spin-1 graviton modes of the Minkowski background. Indeed focussing again on propagation in the $z$ direction, the $x-y$ rotation symmetry gives a higher degeneracy with these 6 modes arranging themselves into a non-degenerate pair, and two degenerate pairs, corresponding to the spin-0 and spin-1 modes respectively. However for propagation in a general direction $\hat{k}$ they are fully non-degenerate, as we see from the example of the characteristic polynomial  above for propagation in the $x$-direction. Thus, unlike the spin-2 modes, we see the spin-1 modes are not simply controlled by the lightcone of some metric, where both polarizations propagate identically. Instead the two polarizations will be governed by different rays, and so the theory is birefringent. Such birefringence has been observed in other modified gravity theories \cite{Garfinkle:2010, Jenks:2023pmk}. 
We also note that, as observed in~\cite{Gruzinov:2011sq,Deser:2012qx,Deser:2013eua}, one of the non-degenerate graviton wavemodes generically will be `faster than light', by which we mean that it propagates outside the metric lightcone.

A further comment is that in this example the physical eigenvalues that split include a pair governed by only the vierbein, from the quadratic factor $\left( 2 \Lambda^2 + \phi (1 - \phi) \right)$ in the 18th order polynomial~\eqref{eq:polyphys}, with the remaining 4 from the quartic factor  depending on both the vierbein (through $\phi$) and the derivative variable $Q_{xyt}$. This however is not a generic splitting in the sense that in the most general situation the wavemode factor $P_{wave}(\Lambda)$ of the characteristic polynomial of $\mat{M}_x$ comprises 6 quadratic factors from the degenerate eigenvectors, and a 6th order polynomial that involves both the vierbein variables and the derivative variables -- thus all the eigenvalues that split to be non-degenerate explicitly depend on these derivative variables. This can be clearly seen in a more complicated analytic example. If we deform the spatial vierbein to 
\be
e_{ij} = \left(\begin{matrix} 
1 & \beta & 0 \\
\beta & 1 & 0 \\
0 & 0 & 1\end{matrix}\right)
\ee
 and include in addition to $Q_{xyt}$ also the derivative variables $Q_{zxx}$, $P_{xy}$ and $P_{z}$, then one indeed finds this structure. Solving the vector and scalar constraints for the background then determines $P_z$, $P_{xy}$ and $Q_{zxx}^2$. The resulting wavemode factor of the characteristic polynomial, $P_{wave}(\Lambda)$, for $\mat{M}_x$ again splits into the 6 degenerate pairs given by the factor $( \Lambda^2 - \phi^2 )^6$ and now a complicated sixth order polynomial which generally does not factorise to a factor that is independent of the derivative variables.

\section{Well-posedness near Minkowski spacetime}
\label{sec:wellposedMink}

Above we have shown that the zero eigenvectors of Minkowski remain zero eigenvectors on a generic background. Six pairs of eigenmodes -- the `wavemode' eigenmodes -- remain degenerate on a generic background and their eigenvalues are controlled by the inverse metric. Also in the special case of interest that $\tilde{e}_{ij} = e_{ij}$ in the background, two pairs of eigenvalues remain degenerate and are controlled only by the background vierbein and the value of $\lambda$.
We now argue that in a generic neighbourhood of Minkowski, 
at a generic point and for a generic wavevector direction $\hat{k}_i$,
these are the \emph{only} degenerate eigenvectors, with all others being non-degenerate and having real eigenvalue. 
However as we will discuss, this argument fails at a generic point for a set of measure zero wavevector directions. It may also fail at special points.

The essence of the argument is to consider the eigenvalues for a small perturbation about Minkowski spacetime and show that the degenerate eigenvalues split apart, leaving only the degeneracies we have shown generically exist, and do so in a manner that they must remain real -- recall eigenvalues can only become complex in conjugate pairs, so it is sufficient to show that as we perturb, the non-degenerate eigenvalues attain unequal real parts. Recall also that eigenvectors must exist for non-degenerate eigenvalues -- it is only in the case of degenerate eigenvalues that they can fail to exist, as now the associated Jordan blocks may have off-diagonal elements, whereas for a single non-degenerate eigenvalue the Jordan block is only one-by-one. 

It is relatively straightforward to compute the perturbation to the eigenvalues as we perturb the background from Minkowski. 
By construction, the variables of our first order system have been chosen to preserve the harmonic Einstein equation's spatial rotational invariance, and further here we will consider a general background, so we may restrict attention to the derivatives in the $x$-direction (so $\hat{k}_i = (1,0,0)$) without loss of generality, and hence focus our attention on $\mat{M}_x$.

We take the background vector $\matvec{u}$ in equation~\eqref{eq:uvec} to be a perturbation of Minkowski as,
\be
\matvec{u} = \matvec{u}^0 + \epsilon \, \delta \matvec{u}
\ee
where $\matvec{u}^0$ has only non-vanishing components $\phi = -1$ and $e_{xx} = e_{yy} = e_{zz} = 1$, and $\delta \matvec{u} = \left( \delta \phi, \delta P_x, \ldots, \delta \tilde{e}_{zx} \right)$. We focus on the $x$-derivatives, and hence on $\mat{M}_x$, which takes the form,
\be
\mat{M}_x = \mat{M}^0_x + \epsilon \mat{M}^1_x + O(\epsilon^2) \; .
\ee
Now conjugating by the same matrix $\mat{S}$ that puts the Minkowski background in Jordan form, ie. that in equation~\eqref{eq:MinkJordan} and given explicitly in Appendix \ref{app:details} then we find,
\be
\label{eq:conj_mat}
\mat{M}^{conj}_x \equiv \mat{S}^{-1} \cdot \mat{M}_x \cdot \mat{S} = \left(
\begin{array}{c|c|c|c|c}
-\mat{1}_9 + \epsilon \mat{M}^{-} & O(\epsilon) & O(\epsilon) & O(\epsilon) & O(\epsilon) \\ \hline
O(\epsilon) & + \mat{1}_9 + \epsilon \mat{M}^{+} & O(\epsilon) & O(\epsilon) &O(\epsilon)  \\ \hline
O(\epsilon)& O(\epsilon)& + \mat{0}_{10} & O(\epsilon) & O(\epsilon) \\  \hline
O(\epsilon)&O(\epsilon) & O(\epsilon)& 
\begin{array}{cccc}
- \lambda & 0 & 0 & 0 \\
0 & - \lambda & 0 & 0 \\
0 & 0 & + \lambda  & 0\\
0 & 0 & 0 & + \lambda \\
\end{array} + O(\epsilon)
&O(\epsilon)
 \\  \hline
O(\epsilon)& O(\epsilon)& O(\epsilon)& O(\epsilon)&
\begin{array}{cc}
- \mu & 0  \\
& + \mu
\end{array}
\end{array}
\right) + O(\epsilon^2) \; . \nl 
\ee
For completeness we give the matrices $\mat{M}^{\pm}$ also in Appendix \ref{app:details}.
Let $\matvec{v}$ be an eigenvector of $\mat{M}^{conj}_x$; then $\mat{S} \cdot \matvec{v}$ is an eigenvector of $\mat{M}_x$ with the same eigenvalue. 
Above we have shown that the zero eigenvectors associated to the 10 by 10 block of zeros on the diagonal remain for generic backgrounds.
As shown above, the 2 by 2 block controlled by $\mu$ is associated to eigenvectors with values $\pm {\mu}$ on a generic background. Further the 4 by 4 block controlled by $\lambda$ gives 4 eigenvectors on a generic background, which are non-degenerate in general but become two degenerate pairs in the special case of interest that $\tilde{e}_{ij} = e_{ij}$.
Thus our focus is on the remaining two 9 by 9 blocks on the diagonal in the top left above, which give eigenvectors with eigenvalues $\pm 1$ in Minkowski.

Let us suppose that we have  eigenvectors $\matvec{\hat{v}}^{\pm}$ of the 9 by 9 matrices $\mat{M}^{\pm}$ that perturb the leftmost diagonal blocks, whose eigenvalues are $\Delta\Lambda^{\pm}$ so that,
\be
\mat{M}^{\pm} \cdot  \matvec{\hat{v}}^{\pm} = \Delta\Lambda^{\pm}  \matvec{\hat{v}}^{\pm} \; .
\ee
Then (following the degenerate perturbation theory procedure that is so familiar from QM) we may construct eigenvectors of the full $\mat{M}^{conj}_x$ as,
\be
\matvec{v} = \left(  \matvec{\hat{v}}^{-} , \matvec{0}_{9} , \matvec{0}_{16} \right) + O(\epsilon) \; , \quad \Lambda = -1 + \epsilon \Delta\Lambda^{-} + O(\epsilon^2)
\ee
and
\be
\matvec{v} = \left(  \matvec{0}_{9} , \matvec{\hat{v}}^{+} ,  \matvec{0}_{16} \right) + O(\epsilon) \; , \quad \Lambda = +1 + \epsilon \Delta\Lambda^{+} + O(\epsilon^2) \; .
\ee
Thus the leading perturbation to the degenerate eigenvalues is given the eigenvalues of $\mat{M}^{\pm}$.
We may straightforwardly compute these eigenvalues and eigenvectors. We find:
\begin{itemize} 
\item
6 degenerate eigenvectors with eigenvalues,
\be
 \Delta\Lambda^{\pm} =  \mp \left( \delta \phi + \delta e_{xx} \mp 2 \delta V_x \right)
\ee 
so that $\Lambda = \pm 1 \mp \epsilon ( \delta \phi + \delta e_{xx} \mp 2 \delta V_x ) + O(\epsilon^2)$ corresponds to the linearization of the  6 degenerate eigenvalues of the wavemode eigenmodes we have discussed above in section~\ref{sec:physmodes}, given by the inverse metric lightcone condition $g^{\mu\nu} \hat{k}_\mu \hat{k}_\nu = 0$.
\item
1 eigenvector whose eigenvalue differs by a factor of $3/4$ from these degenerate ones,
\be
\Delta\Lambda^{\pm} =  \mp \frac{3}{4} \left( \delta \phi + \delta e_{xx} \mp 2 \delta V_x \right)
\ee
\item
a pair of eigenvalues given as roots of a quadratic;
\be
\Delta\Lambda^{\pm} &=&  \mp \frac{7}{8} \left( \delta \phi + \delta e_{xx} \mp 2 \delta V_x \right) \\
& \pm& \frac{1}{8}  \sqrt{  \left(
 \left(  \delta \phi + \delta e_{xx} \mp 2 \delta V_x \right)^2 
+ \frac{16}{3 m^2} \left(   2 \delta Q_{xyt} \mp \delta Q_{xyx} + \delta P_{xy} \mp \delta P_{y}\right)^2
 + \frac{16}{3 m^2} \left(  \delta Q_{zxt} \mp \delta Q_{zxx} - \delta P_{zx} \pm \delta P_{z}  \right)^2
 \right) }  \nonumber
\ee
which importantly is always real since the square root is taken on a sum of positive terms.
\end{itemize}
Thus for a generic perturbation of Minkowski, and at a generic point, here meaning that, $\delta \phi + \delta e_{xx} \mp 2 \delta V_x$ is non-vanishing, and likewise  either $2 \delta Q_{xyt} \mp \delta Q_{xyx} + \delta P_{xy} \mp \delta P_{y}$ or $\delta Q_{zxt} \mp \delta Q_{zxx} - \delta P_{zx} \pm \delta P_{z}$ or both, 
then we see the expected 6 degenerate eigenvalues and 3 non-degenerate \emph{real} eigenvalues at leading order $O(\epsilon)$  in the perturbation. 
One might be concerned that these eigenvectors have only been constructed to $O(\epsilon)$ and hence could be obstructed from existing at higher order in perturbation theory, or  become complex.
However from above we know that these 6 degenerate eigenvalues indeed continue to be real and correspond to linearly independent eigenvectors in a generic neighbourhood of Minkowski. Then since the remaining 3 eigenvalues become non-degenerate at $O(\epsilon)$, and in particular, their real parts all differ at this order, then they must remain real since complex eigenvalues must come in conjugate pairs with equal real part. Furthermore distinct real eigenvalues must be associated to eigenvectors.

An important subtlety is that we have restricted ourselves to considering the $x$ direction, so studying $\mat{M}_x$. To obtain the expression for a general direction one simply should perform a rotation on the metric, momentum and derivative variables. Since we have two angles to specify a direction, it may be that for particular directions either the combination $\delta \phi + \delta e_{xx} \mp 2 \delta V_x$ or the pair $2 \delta Q_{xyt} \mp \delta Q_{xyx} + \delta P_{xy} \mp \delta P_{y}$ or $\delta Q_{zxt} \mp \delta Q_{zxx} - \delta P_{zx} \pm \delta P_{z}$ vanish for one of the sign choices. 
Such additional degeneracy would
generically occur for a set of codimension one in the space of wavevector directions $\hat{k}$ at a point. Thus we cannot rule out eigenvalues of  $\mat{M}[\hat{k}_i]$ becoming defective at some point for a set of a zero measure of wavevector directions, and for this reason 
we cannot prove well-posedness of our formulation for backgrounds in the neighborhood of Minkowski.
Likewise there may exist special points for a given background, where there are additional degeneracies.

Hence for a generic point and direction $\hat{k}_i$ and a generic neighbourhood of Minkowski, a complete set of eigenvectors exist, and the degeneracies are only the ones where we have shown eigenvectors generically exist; the 10 zero eigenvectors, the 6 pairs governed by the inverse metric, and for the special case of interest, $\tilde{e}_{ij} = e_{ij}$, the two pairs governed by the vierbein and $\lambda$. However we are careful to note that we have not yet managed to control `accidental degeneracies' for particular non-generic directions $\hat{k}_i$ and the argument does not constitute a proof of well-posedness. 
In the case of the analytic example of section~\ref{sec:analyticexample} we saw that for particular wavevector directions the eigenvectors became degenerate but in that case a complete set still existed. For the Minkowski background all eigenvectors are degenerate and the theory is well-posed. 
Thus while we cannot prove well-posedness, this certainly does not imply the formulation is ill-posed. It simply means that our argument cannot demonstrate it is well-posed.

\section{Behaviour far from Minkowski}

%
What happens far from Minkowski spacetime? From the previous discussion we expect the matrix $\mat{A}$ to be generically invertible, and we know that the characteristic polynomial of $\mat{M}_x$ takes the form,
\be
\label{eq:charpoly}
P(\Lambda) &=& \Lambda^{10} ( \Lambda^2 - \mu^2) 
\left( \left. \Lambda^2 - \frac{\lambda^2}{1 - \lambda^2}  \hat{k}_\mu \hat{k}_\nu (E^{-1})^{\mu\nu} \right|_{e_{yz} \to \tilde{e}_{yz}, e_{zx} \to \tilde{e}_{zx}} \right) 
\left( \left. \Lambda^2 - \frac{\lambda^2}{1 - \lambda^2}  \hat{k}_\mu \hat{k}_\nu (E^{-1})^{\mu\nu} \right|_{ e_{xy} \to \tilde{e}_{xy}, e_{zx} \to \tilde{e}_{zx}} \right) \nl
&& \qquad \qquad \times \left( g^{\mu\nu} \hat{k}_\mu \hat{k}_\nu \right)^6
P_{phys}(\Lambda)  \nl
\ee
where, as above, $\hat{k}_\mu = (\Lambda, 1,0,0)$ and we see the 10 zero eigenvalues, the two with eigenvalues $\pm \mu$ and the 4 that continue from the pair of eigenvalues $\pm \lambda$ for the Minkowski background (as discussed in section~\ref{sec:lambdamodes}). We also see the 6 degenerate pairs of eigenvalues controlled by the inverse metric, and then the remainder, $P_{phys}$, is a 6th order polynomial controlling the non-degenerate eigenvalues that continue from the values $\pm 1$ on Minkowski, and correspond to the spin-0 and spin-1 wavemodes  of the graviton on the Minkowski background.

For well-posedness far from Minkowski we will require that the metric $g_{\mu\nu}$ is Lorentzian to ensure that the degenerate wavemode eigenvectors all have real eigenvalues -- this is simply the same requirement as for GR. Likewise we also require that the roots of the quadratic $\left(\frac{\lambda^2}{1 - \lambda^2} (E^{-1})^{\mu\nu} \hat{k}_\mu \hat{k}_\nu - \Lambda^2\right)$ remain real, where for derivatives in a general direction $\hat{k}_i$ then $\hat{k}_\mu = (\Lambda,  \hat{k}_i )$. Writing the vector $n = \partial/\partial t$ associated to our time variable, then we may write this as the condition that,
\be
\label{eq:Gfmetric}
G_{(\lambda)}^{\mu\nu} \hat{k}_\mu \hat{k}_\nu  = 0 \; , \quad  G_{(\lambda)}^{\mu\nu} =  (E^{-1})^{\mu\nu}  - \frac{(1 - \lambda^2)}{\lambda^2} n^\mu n^\nu
\ee
has real solutions for $\hat{k}_\mu$ for any direction $\hat{k}_i$. This in turn is the condition that $G_{(\lambda)}^{\mu\nu} $ defined above is a Lorentzian inverse metric. We note that this condition depends explicitly on our choice of time coordinate. 
In the analysis above we are free to choose any positive $\lambda$, provided we don't pick $\lambda = 1$. Hence we may regard this condition above as one that constrains the allowed range for $\lambda$. If one performed a numerical simulation, it would even be possible to adjust the value of $\lambda$ as the evolution proceeds in order to ensure this condition is met.

The 6th order polynomial $P_{phys}(\Lambda)$ takes a complicated form and we have not yet found a covariant formulation of it -- presumably it can be written neatly as $\hat{k}_\alpha \hat{k}_\beta \hat{k}_\sigma \hat{k}_\rho \hat{k}_\mu \hat{k}_\nu P^{\alpha\beta\sigma\rho\mu\nu}$ where the 6 index tensor $P^{\alpha\beta\sigma\rho\mu\nu}$ has a relatively simple form in terms of the vierbein and its inverse, the reference metric, and the derivative variables. 
What we can say  is that the 6 roots of this polynomial will generically not be degenerate with any of the other roots of $P(\Lambda)$ and hence there must be corresponding eigenvectors. However it may be that these eigenvalues fail to be real. In this case our formulation becomes ill-posed, but then the dRGT wavemodes will have complex wavevectors in the short wavelength limit. This is then a feature of dRGT rather than our formulation. 
It would be interesting to give a general analysis of the characteristic structure of dRGT but it is beyond the scope of this paper and we leave it for future work.

\section{Constraints on initial data for dRGT solutions}
\label{sec:initialdata}

We have expanded the minimal dRGT theory firstly to the harmonic formulation, and then further to a first order formulation. In both steps we introduce additional variables and hence initial data. Solutions of these expanded systems only obey the original dRGT equations if the initial data obeys specific conditions. As we have seen, solutions to the harmonic equations only correspond to those of dRGT if $\xi_\mu$ and $\dot{\xi}_{\mu} = 0$ in the initial data. For the first order formulation we expand the variables to include those representing derivatives of the vierbein -- however then in the initial data they must be set accordingly.

The physical dRGT theory possesses the $5$ graviton second-order degrees of freedom, which in our first-order formulation translates to $10$ variables whose values we can freely specified at the start of the evolution. Since in our expanded first order system we have $34$ variables in total, there should consequently be $34-10 =24$ independent constraints on the initial data if it is to evolve to a dRGT solution. Explicitly these are as follows:
\begin{itemize}
    \item The vanishing of the vector constraint $\xi_\mu$ and its time derivative ($8$ conditions).
    \item The scalar constraint $\mathcal{S}$ (one condition).
    \item The equality of the auxiliary variables $\tilde{e}_{ij}$ to their physical counterparts ($3$ conditions).
    \item The definitions of the $12$ `$Q$' variables, that is, $Q_{ij\mu}=\p_i E_{j\mu} - \p_j E_{i\mu}$. Note that this ensures the variables $J$ and $I_i$, used in our well-posed formulation, are indeed equal to zero ($12$ conditions).
\end{itemize}
It is straightforward to see that the conditions above are preserved under time evolution. Let us therefore assume they  hold at some time $t_0$. Then,
\begin{itemize}
    \item The vector constraint $\xi_\mu$ obeys the wavelike equation \eqref{eq:harmonicBianchi} so, as discussed previously, imposing that it and its first time derivative vanish at some time implies that it will remain zero at later times.
    \item The equation $\p_t \mathcal{S}=0$ is imposed directly, so $S$ will remain zero.
    \item Consider the time derivative of the difference $\tilde{e}_{ij}-e_{ij}$. We have
    \be
        \p_t\left(\tilde{e}_{ij}-e_{ij}\right) &=& \left(P_{ij}+Q_{ijt}+\p_j V_i\right) - \left(P_{ij}+\p_i V_j\right) \nl
        &=& Q_{ijt} -\left(\p_iV_j-\p_jV_i\right)
    \ee
    which vanishes, since we have assumed the definition of $Q_{ijt}$ is obeyed at $t_0$. Hence $\tilde{e}_{ij}$ remains equal to $e_{ij}$.
    \item Finally, since we assumed the definitions of $Q$ are obeyed, we also have $J = I_i = 0$ for $i=x,y,z$. We also imposed $\xi_\mu=0$ and $e_{ij} = \tilde{e}_{ij}$ at $t_0$, so we can ignore the modification of the evolution equations for $V_i$. Therefore we can write, for example
    \be 
        \p_t\left(Q_{xyy}-(\p_x e_{yy}-\p_y e_{zx})\right) &=& \p_x P_{yy}-\p_y P_{xy} -\p_x(\p_te_{yy})+\p_y (\p_te_{xy}) \nl
        &=& \p_x(P_{yy}-\p_t e_{yy})-\p_y(P_{xy}-\p_t e_{xy}) \nl
        &=& -\p_x\p_y V_y + \p_y\p_x V_y = 0
    \ee
    or
    \be 
        \p_t\left(Q_{xyz}-(\p_x e_{yz}-\p_y e_{xy})\right) &=&
        \p_z Q_{xyt}+\p_x Q_{yzt} +\p_x P_{yz} - \p_y P_{zx} - \p_x\left(P_{yz}+\p_y V_z\right) +\p_y \left(P_{zx} +\p_zV_x\right) \nl
        &=& \p_zQ_{xyt} + \p_x Q_{yzt} + \p_y Q_{zxt} = 0
    \ee
    with analogous results for the remaining $Q_{ij\mu}$ variables. Hence these constraints  are preserved due to the definitions of the `$Q$' variables.
\end{itemize}

\section{Conclusion and future directions}

Here we have considered the minimal dRGT theory as a classical p.d.e. system and asked whether it is well-posed or not. It need not be well-posed, as it is regarded as an effective field theory, 
and well-posedness is decided precisely by the short distance behaviour of the theory where high dimension operators are expected to become important.
We have given a first order formulation of a natural harmonic formulation that we have proved is well-posed 
on the Minkowski vacuum spacetime.
Our discussion of
  well-posedness near Minkowski spacetime falls short of a rigorous proof. We have shown the requisite properties for well-posedness hold for propagation in a generic direction $\hat{k}_i$ on a generic background. However it may be that non-generic backgrounds yield additional degeneracies; further particular directions $\hat{k}_i$ may have enhanced degeneracies. We do not believe such extra degeneracy really threatens well-posedness for backgrounds near Minkowski spacetime, but without demonstrating that $\mat{M}[ \hat{k}_i ]$ has its full complement of eigenvectors in these non-generic situations we cannot say we have formally proved well-posedness. We emphasize however that we believe the system is well-posed, and we hope to complete the proof in future work.

We have discussed well-posedness for backgrounds that are a large deformations away from Minkowski spacetime. A lack of well-posedness may then result from;
\begin{itemize}
\item the metric $g_{\mu\nu}$ failing to be Lorentzian, in which case the degrees of freedom associated to the spin-2 graviton modes (and also the constraint violating modes) would not be wavelike,
\item the 6 roots of $P_{wave}(\Lambda)$ defined in equation~\eqref{eq:charpoly}, now for a general direction $\hat{k}_i$ failing to be real; then the degrees of freedom associated to the spin-0 and spin-1 graviton modes would fail to be wavelike,
\item the roots of $P_{wave}(\Lambda)$ remaining real, but failing to have a corresponding full set of eigenvectors at particular spacetime points or directions $\hat{k}_i$,
\item the inverse metric $G^{\mu\nu}_{(\lambda)}$ defined in~\eqref{eq:Gfmetric} failing to be Lorentzian; we emphasize that this is not a covariant condition, and depends on the choice of time coordinate. Further it can be thought of as constraining the value of $\lambda$, which we are free to choose.
\end{itemize}
If the first two are the cause then this implies that the ill-posedness is  a property of the dRGT dynamics rather than our particular formulation. In the last case this would be due to our formulation, and in particular may be a result of the choice of coordinates -- we note that even having taken coordinates so that the reference metric is $\eta_{\mu\nu}$ there is the residual Poincare coordinate freedom. Of course another obstruction to time evolution is that singularities may form in the components of the vierbein, or its derivatives. This may be due to a physical singularity forming, or alternatively might be due to an inappropriate choice of coordinate system. However  we emphasize that unlike for GR where we may have a seemingly singular metric become smooth after an appropriate singular diffeomorphism, one cannot so straightforwardly have this situation in the dRGT theory since a singular diffeomorphism would generally render the reference metric to be singular even if it did resolve a singularity in the metric.

We have formulated the harmonic Einstein equations, and our first order formalism, in the unitary gauge where we have chosen the reference metric to be the Minkowski metric in usual coordinates. Since we have restricted our full analysis to small deformations of the Minkowski vacuum, this is not a serious restriction. However, if one were to study dynamics involving larger departures from Minkowski, such as for gravitational collapse, or such as are expected for a working Vainshtein mechanism, one might wish to build a first order formulation based on a different coordinate system -- for example, one adapted to the formation of a horizon.
The harmonic Einstein equations in~\eqref{eq:harmK} may be simply transformed to general coordinates by transforming the reference metric $\eta_{\mu\nu} \to f_{\mu\nu}$ by a diffeomorphism, and further replacing the partial derivatives in these expressions with the covariant derivative, $\nabla^{(f)}$, of this new reference metric. It is then natural to change the first order variables too by defining them via,
\be
K_{\alpha\beta\mu} = \nabla^{(f)}_{\alpha} E_{\beta\mu} - \nabla^{(f)}_{\beta} E_{\alpha\mu} \; ,
\ee
where we again take derivative variables as $P_i = K_{t i t}$, $P_{ij} = K_{t i j}$, and $Q_{ij\mu} = K_{ij\mu}$.
We then expect a first order formulation, using the same variables as we have chosen here in unitary gauge, that will  yield a well-posed system for suitable choices of coordinates. We will explore this further in future work, the expectation being that this  might allow ill-posedness or failure in evolution associated to the coordinate dependence to be cured; for example $G^{\mu\nu}_{(\lambda)}$ failing to be Lorentzian due to the choice of time coordinate.

We have not analysed the full characteristic structure of the theory far from Minkowski. This involves understanding the 6 eigenvalues for a general background that correspond to the spin-0 and spin-1 wavemodes in the Minkowski vacuum. This appears to be a difficult task which we leave  for future work.
Another future direction which we are pursuing is to give a similar formulation which includes the next-to-minimal mass term. A similar first order dynamical formulation was detailed in~\cite{deRham:2023ngf}, although not for a harmonic formulation. 
We have based our first order formulation on the harmonic formulation of dRGT. It would be interesting to understand if we may obtain a well-posed first order formulation of the dRGT theory itself, rather than first extending it to this harmonic form.

Of course whether this minimal theory has a  good Vainshtein screening mechanism, and hence whether it can reproduce GR-like phenomenology on small scales, is a complicated question. It is known this cannot happen in spherical symmetry, but it may be possible to achieve in generic settings~\cite{Renaux-Petel:2014pja}. This underscores the importance of having a well-posed dynamical formulation of these dRGT theories so that their general dynamics can be revealed using numerical simulation.

While our focus here was on well-posedness of the theory, we have also revealed some interesting results on the characteristics of the dRGT minimal theory. In particular we have shown that the two degrees of freedom associated to the spin-2 modes on Minkowski are always governed by the inverse metric light cone. 
We expect that this is an important feature for potential agreement with gravitational wave phenomenology, and we would want this to persist also when the non-minimal mass term is added if that theory is to be 
phenomenologically viable. 
We have found that the remaining 3 degrees of freedom that on flat spacetime comprise the spin-1 and spin-0 wavemodes have a considerably more complicated structure. Indeed on a general background the degeneracy of all these modes is broken, so that they all have different characteristics. This implies a birefringence in propagation for modes that continue from the spin-1 graviton so that the two polarizations propagate differently through a general background.

\subsection*{Acknowledgments}
We thank Claudia de Rham, Harvey Reall and Andrew Tolley for very helpful discussions.
This work is supported by STFC Consolidated Grant ST/T000791/1. JK is funded by an STFC studentship.
We are very grateful to the referee of this paper for their careful reading and valuable comments.

\appendix 

\section{Details for the Minkowski background and its perturbation}
\label{app:details}
\raggedbottom

The matrices $\mat{M}_x$ and $\mat{S}$ from Eq. \eqref{eq:firstorderwellpos} are given by
\begingroup
\setlength{\arraycolsep}{2pt}
\begin{equation*}
\mat{M}_x= \small{\left(
\begin{array}{cccccccccccccccccccccccccccccccccc}
 0 & 0 & 0 & 0 & 0 & 0 & 0 & 0 & 0 & 0 & 1 & 0 & 0 & 0 & 0 & 0 & 0 & 0 & 0 & 0 & 0 & 0 & 0 & 0 & 0 & 0 & 0 & 0 & 0 & 0 & 0 & 0 & 0 & 0 \\
 0 & 0 & 0 & 0 & -1 & -1 & -1 & 0 & 0 & 0 & 0 & 0 & 0 & 0 & 0 & 0 & 0 & 0 & 0 & 0 & 0 & 0 & 0 & 0 & 0 & 0 & 0 & 0 & 0 & 0 & 0 & 0 & 0 & 0 \\
 0 & 0 & 0 & 0 & 0 & 0 & 0 & 0 & 0 & 0 & 0 & 0 & 0 & 0 & 0 & 0 & 0 & 0 & 0 & 0 & 0 & 0 & 0 & 0 & 0 & 1 & 0 & 0 & 0 & 0 & 0 & 0 & 0 & 0 \\
 0 & 0 & 0 & 0 & 0 & 0 & 0 & 0 & 0 & 0 & 0 & 0 & 0 & 0 & 0 & 0 & 0 & 0 & 0 & 0 & 0 & 0 & 0 & 0 & 0 & 0 & 0 & -1 & 0 & 0 & 0 & 0 & 0 & 0 \\
 0 & -1 & 0 & 0 & 0 & 0 & 0 & 0 & 0 & 0 & 0 & 0 & 0 & 0 & 0 & 0 & 0 & 0 & 0 & -1 & 0 & 0 & 0 & 0 & 1 & 0 & 0 & 0 & 0 & 0 & 0 & 0 & 0 & 0 \\
 0 & 0 & 0 & 0 & 0 & 0 & 0 & 0 & 0 & 0 & 0 & 0 & 0 & 0 & 0 & 0 & 0 & 0 & 0 & 1 & 0 & 0 & 0 & 0 & 0 & 0 & 0 & 0 & 0 & 0 & 0 & 0 & 0 & 0 \\
 0 & 0 & 0 & 0 & 0 & 0 & 0 & 0 & 0 & 0 & 0 & 0 & 0 & 0 & 0 & 0 & 0 & 0 & 0 & 0 & 0 & 0 & 0 & 0 & -1 & 0 & 0 & 0 & 0 & 0 & 0 & 0 & 0 & 0 \\
 0 & 0 & -1 & 0 & 0 & 0 & 0 & 0 & 0 & 0 & 0 & 0 & 0 & 0 & 0 & 0 & 0 & 0 & 0 & 0 & 1 & -\frac{\mu ^2+1}{2}  & 0 & 0 & 0 & 0 & 0 & 0 & 0 & 0 & 0 & 0 & 0 & 0 \\
 0 & 0 & 0 & 0 & 0 & 0 & 0 & 0 & 0 & 0 & 0 & 0 & 0 & 0 & 0 & 0 & 0 & 0 & 0 & 0 & 0 & 0 & 0 & 0 & 0 & 0 & 0 & 0 & \frac{1}{2} & -\frac{\mu ^2}{2} & -\frac{1}{2} & 0 & 0 & 0
   \\
 0 & 0 & 0 & 0 & 0 & 0 & 0 & 0 & 0 & 0 & 0 & 0 & 0 & 0 & 0 & 0 & 0 & 0 & 0 & 0 & 0 & 0 & \frac{1-\mu ^2}{2}  & -1 & 0 & 0 & 0 & 0 & 0 & 0 & 0 & 0 & 0 & 0 \\
 1 & 0 & 0 & 0 & 0 & 0 & 0 & 0 & 0 & 0 & 0 & 0 & 0 & 0 & -\lambda ^2 & -\lambda ^2 & 0 & 0 & 0 & 0 & 0 & 0 & 0 & 0 & 0 & 0 & 0 & 0 & 0 & 0 & 0 & 0 & 0 & 0 \\
 0 & 0 & 0 & 0 & 0 & 0 & 0 & 0 & 0 & 0 & 0 & 0 & 0 & 0 & 0 & 0 & \lambda ^2 & 0 & 0 & 0 & 0 & 0 & 0 & 0 & 0 & 0 & 0 & 0 & 0 & 0 & 0 & 0 & 0 & 0 \\
 0 & 0 & 0 & 0 & 0 & 0 & 0 & 0 & 0 & 0 & 0 & 0 & 0 & 0 & 0 & 0 & 0 & 0 & 0 & 0 & 0 & 0 & 0 & 0 & 0 & 0 & 0 & 0 & 0 & 0 & 0 & 0 & 0 & \lambda ^2 \\
 0 & 0 & 0 & 0 & 0 & 0 & 0 & 0 & 0 & 0 & 1 & 0 & 0 & 0 & 0 & 0 & 0 & 0 & 0 & 0 & 0 & 0 & 0 & 0 & 0 & 0 & 0 & 0 & 0 & 0 & 0 & 0 & 0 & 0 \\
 0 & 0 & 0 & 0 & 0 & 0 & 0 & 0 & 0 & 0 & 0 & 0 & 0 & 0 & 0 & 0 & 0 & 0 & 0 & 0 & 0 & 0 & 0 & 0 & 0 & 0 & 0 & 0 & 0 & 0 & 0 & 0 & 0 & 0 \\
 0 & 0 & 0 & 0 & 0 & 0 & 0 & 0 & 0 & 0 & 0 & 0 & 0 & 0 & 0 & 0 & 0 & 0 & 0 & 0 & 0 & 0 & 0 & 0 & 0 & 0 & 0 & 0 & 0 & 0 & 0 & 0 & 0 & 0 \\
 0 & 0 & 0 & 0 & 0 & 0 & 0 & 0 & 0 & 0 & 0 & 1 & 0 & 0 & 0 & 0 & 0 & 0 & 0 & 0 & 0 & 0 & 0 & 0 & 0 & 0 & 0 & 0 & 0 & 0 & 0 & 0 & 0 & 0 \\
 0 & 0 & 0 & 0 & 0 & 0 & 0 & 0 & 0 & 0 & 0 & 0 & 0 & 0 & 0 & 0 & 0 & 0 & 0 & 0 & 0 & 0 & 0 & 0 & 0 & 0 & 0 & 0 & 0 & 0 & 0 & 0 & 0 & 0 \\
 0 & 0 & 0 & 0 & 0 & 0 & 0 & 0 & 0 & 0 & 0 & 0 & 0 & 0 & 0 & 0 & 0 & 0 & 0 & 0 & 0 & 0 & 0 & 0 & 0 & 0 & 0 & 0 & 0 & 0 & 0 & 0 & 0 & 0 \\
 0 & 0 & 0 & 0 & 0 & 1 & 0 & 0 & 0 & 0 & 0 & 0 & 0 & 0 & 0 & 0 & 0 & 0 & 0 & 0 & 0 & 0 & 0 & 0 & 0 & 0 & 0 & 0 & 0 & 0 & 0 & 0 & 0 & 0 \\
 0 & 0 & 0 & 0 & 0 & 0 & 0 & 1 & 0 & 0 & 0 & 0 & 0 & 0 & 0 & 0 & 0 & 0 & 0 & 0 & 0 & 0 & 0 & 0 & 0 & 1 & 0 & 0 & 0 & 0 & 0 & 0 & 0 & 0 \\
 0 & 0 & 0 & 0 & 0 & 0 & 0 & 0 & 0 & 0 & 0 & 0 & 0 & 0 & 0 & 0 & 0 & 0 & 0 & 0 & 0 & 0 & 0 & 0 & 0 & 0 & 0 & 0 & 0 & 0 & 0 & 0 & 0 & 0 \\
 0 & 0 & 0 & 0 & 0 & 0 & 0 & 0 & 0 & 0 & 0 & 0 & 0 & 0 & 0 & 0 & 0 & 0 & 0 & 0 & 0 & 0 & 0 & 0 & 0 & 0 & 0 & 0 & 0 & 0 & 0 & 0 & 0 & 0 \\
 0 & 0 & 0 & 0 & 0 & 0 & 0 & 0 & 0 & -1 & 0 & 0 & 0 & 0 & 0 & 0 & 0 & 0 & 0 & 0 & 0 & 0 & 0 & 0 & 0 & 0 & 0 & 0 & 0 & 0 & 0 & 0 & 0 & 0 \\
 0 & 0 & 0 & 0 & 0 & 0 & -1 & 0 & 0 & 0 & 0 & 0 & 0 & 0 & 0 & 0 & 0 & 0 & 0 & 0 & 0 & 0 & 0 & 0 & 0 & 0 & 0 & 0 & 0 & 0 & 0 & 0 & 0 & 0 \\
 0 & 0 & 1 & 0 & 0 & 0 & 0 & 0 & 0 & 0 & 0 & 0 & 0 & 0 & 0 & 0 & 0 & 0 & 0 & 0 & 0 & \mu ^2 & 0 & 0 & 0 & 0 & 0 & 0 & 0 & 0 & 0 & 0 & 0 & 0 \\
 0 & 0 & 0 & 0 & 0 & 0 & 0 & 0 & 0 & 0 & 0 & 0 & 0 & 0 & 0 & 0 & 0 & 0 & 0 & 0 & 0 & 0 & 0 & 0 & 0 & 0 & 0 & 0 & 0 & \mu ^2 & 0 & 0 & 0 & 0 \\
 0 & 0 & 0 & -1 & 0 & 0 & 0 & 0 & 0 & 0 & 0 & 0 & 0 & 0 & 0 & 0 & 0 & 0 & 0 & 0 & 0 & 0 & \mu ^2 & 0 & 0 & 0 & 0 & 0 & 0 & 0 & 0 & 0 & 0 & 0 \\
 0 & 0 & 0 & 0 & 0 & 0 & 0 & 0 & 1 & 0 & 0 & 0 & 0 & 0 & 0 & 0 & 0 & 0 & 0 & 0 & 0 & 0 & 0 & 0 & 0 & 0 & 1 & 0 & 0 & 0 & 0 & 0 & 0 & 0 \\
 0 & 0 & 0 & 0 & 0 & 0 & 0 & 0 & 0 & 0 & 0 & 0 & 0 & 0 & 0 & 0 & 0 & 0 & 0 & 0 & 0 & 0 & 0 & 0 & 0 & 0 & 1 & 0 & 0 & 0 & 0 & 0 & 0 & 0 \\
 0 & 0 & 0 & 0 & 0 & 0 & 0 & 0 & -1 & 0 & 0 & 0 & 0 & 0 & 0 & 0 & 0 & 0 & 0 & 0 & 0 & 0 & 0 & 0 & 0 & 0 & 0 & 0 & 0 & 0 & 0 & 0 & 0 & 0 \\
 0 & 0 & 0 & 0 & 0 & 0 & 0 & 0 & 0 & 0 & 0 & 0 & 0 & 0 & 0 & 0 & 0 & 0 & 0 & 0 & 0 & 0 & 0 & 0 & 0 & 0 & 0 & 0 & 0 & 0 & 0 & 0 & 0 & 0 \\
 0 & 0 & 0 & 0 & 0 & 0 & 0 & 0 & 0 & 0 & 0 & 0 & 0 & 0 & 0 & 0 & 0 & 0 & 0 & 0 & 0 & 0 & 0 & 0 & 0 & 0 & 0 & 0 & 0 & 0 & 0 & 0 & 0 & 0 \\
 0 & 0 & 0 & 0 & 0 & 0 & 0 & 0 & 0 & 0 & 0 & 0 & 1 & 0 & 0 & 0 & 0 & 0 & 0 & 0 & 0 & 0 & 0 & 0 & 0 & 0 & 0 & 0 & 0 & 0 & 0 & 0 & 0 & 0 \\
\end{array}
\right)}
\end{equation*}
\endgroup

\begingroup
\setlength{\arraycolsep}{2pt}
\begin{equation*}
\mat{S} = \small{\left(
\begin{array}{cccccccccccccccccccccccccccccccccc}
 0 & 0 & 0 & 0 & 0 & 0 & 0 & 1 & 0 & 0 & 0 & 0 & 0 & 0 & 0 & 0 & \lambda ^2 & \lambda ^2 & 0 & 0 & 0 & 0 & 0 & 0 & 0 & 0 & 1 & 0 & 0 & 0 & 0 & 0 & 0 & 0 \\
 0 & 0 & 0 & 1 & 0 & 0 & -1 & 0 & 1 & 0 & 0 & 0 & 0 & 0 & 0 & 0 & 0 & 0 & 0 & 0 & 0 & 0 & 1 & 0 & 0 & -1 & 0 & -1 & 0 & 0 & 0 & 0 & 0 & 0 \\
 0 & 0 & -1 & 0 & 0 & 0 & 0 & 0 & 0 & 0 & 0 & 0 & 0 & -\mu ^2 & 0 & 0 & 0 & 0 & 0 & 0 & 0 & 1 & 0 & 0 & 0 & 0 & 0 & 0 & 0 & 0 & 0 & 0 & 0 & 0 \\
 0 & 1 & 0 & 0 & 0 & 0 & 0 & 0 & 0 & 0 & 0 & 0 & -\frac{2 \mu ^2}{\mu ^2-1} & 0 & 0 & 0 & 0 & 0 & 0 & 0 & -1 & 0 & 0 & 0 & 0 & 0 & 0 & 0 & 0 & 0 & 0 & 0 & 0 & 0 \\
 0 & 0 & 0 & 0 & 0 & 0 & 0 & 0 & 1 & 0 & 0 & 0 & 0 & 0 & 0 & 0 & 0 & 0 & 0 & 0 & 0 & 0 & 0 & 0 & 0 & 0 & 0 & 1 & 0 & 0 & 0 & 0 & 0 & 0 \\
 0 & 0 & 0 & 0 & 0 & 0 & -1 & 0 & 0 & 0 & 0 & 0 & 0 & 0 & 0 & 0 & 0 & 0 & 0 & 0 & 0 & 0 & 0 & 0 & 0 & 1 & 0 & 0 & 0 & 0 & 0 & 0 & 0 & 0 \\
 0 & 0 & 0 & 1 & 0 & 0 & 0 & 0 & 0 & 0 & 0 & 0 & 0 & 0 & 0 & 0 & 0 & 0 & 0 & 0 & 0 & 0 & -1 & 0 & 0 & 0 & 0 & 0 & 0 & 0 & 0 & 0 & 0 & 0 \\
 0 & 0 & -1 & 0 & 0 & -1 & 0 & 0 & 0 & 0 & 0 & 0 & 0 & 0 & 0 & 0 & 0 & 0 & 0 & 0 & 0 & -1 & 0 & 0 & 1 & 0 & 0 & 0 & 0 & 0 & 0 & 0 & 0 & 0 \\
 1 & 0 & 0 & 0 & 0 & 0 & 0 & 0 & 0 & 0 & 0 & 0 & 0 & 0 & 0 & 0 & 0 & 0 & 0 & -1 & 0 & 0 & 0 & 0 & 0 & 0 & 0 & 0 & 0 & 0 & 0 & 0 & \mu  & -\mu  \\
 0 & 0 & 0 & 0 & 1 & 0 & 0 & 0 & 0 & 0 & 0 & 0 & 0 & 0 & 0 & 0 & 0 & 0 & 0 & 0 & 0 & 0 & 0 & -1 & 0 & 0 & 0 & 0 & 0 & 0 & 0 & 0 & 0 & 0 \\
 0 & 0 & 0 & 0 & 0 & 0 & 0 & -1 & 0 & 0 & 0 & 0 & 0 & 0 & 0 & 0 & 0 & 0 & 0 & 0 & 0 & 0 & 0 & 0 & 0 & 0 & 1 & 0 & 0 & 0 & 0 & 0 & 0 & 0 \\
 0 & 0 & 0 & 0 & 0 & 0 & 0 & 0 & 0 & 0 & 0 & 0 & 0 & 0 & 0 & 0 & 0 & 0 & 0 & 0 & 0 & 0 & 0 & 0 & 0 & 0 & 0 & 0 & 0 & -\lambda  & 0 & \lambda  & 0 & 0 \\
 0 & 0 & 0 & 0 & 0 & 0 & 0 & 0 & 0 & 0 & 0 & 0 & 0 & 0 & 0 & 0 & 0 & 0 & 0 & 0 & 0 & 0 & 0 & 0 & 0 & 0 & 0 & 0 & -\lambda  & 0 & \lambda  & 0 & 0 & 0 \\
 0 & 0 & 0 & 0 & 0 & 0 & 0 & 1 & 0 & 0 & 0 & 0 & 0 & 0 & 0 & 0 & 0 & 0 & 1 & 0 & 0 & 0 & 0 & 0 & 0 & 0 & 1 & 0 & 0 & 0 & 0 & 0 & 0 & 0 \\
 0 & 0 & 0 & 0 & 0 & 0 & 0 & 0 & 0 & 0 & 0 & 0 & 0 & 0 & 0 & 0 & 0 & 1 & 0 & 0 & 0 & 0 & 0 & 0 & 0 & 0 & 0 & 0 & 0 & 0 & 0 & 0 & 0 & 0 \\
 0 & 0 & 0 & 0 & 0 & 0 & 0 & 0 & 0 & 0 & 0 & 0 & 0 & 0 & 0 & 0 & 1 & 0 & 0 & 0 & 0 & 0 & 0 & 0 & 0 & 0 & 0 & 0 & 0 & 0 & 0 & 0 & 0 & 0 \\
 0 & 0 & 0 & 0 & 0 & 0 & 0 & 0 & 0 & 0 & 0 & 0 & 0 & 0 & 0 & 0 & 0 & 0 & 0 & 0 & 0 & 0 & 0 & 0 & 0 & 0 & 0 & 0 & 0 & 1 & 0 & 1 & 0 & 0 \\
 0 & 0 & 0 & 0 & 0 & 0 & 0 & 0 & 0 & 0 & 0 & 0 & 0 & 0 & 0 & 1 & 0 & 0 & 0 & 0 & 0 & 0 & 0 & 0 & 0 & 0 & 0 & 0 & 0 & 0 & 0 & 0 & 0 & 0 \\
 0 & 0 & 0 & 0 & 0 & 0 & 0 & 0 & 0 & 0 & 0 & 0 & 0 & 0 & 1 & 0 & 0 & 0 & 0 & 0 & 0 & 0 & 0 & 0 & 0 & 0 & 0 & 0 & 0 & 0 & 0 & 0 & 0 & 0 \\
 0 & 0 & 0 & 0 & 0 & 0 & 1 & 0 & 0 & 0 & 0 & 0 & 0 & 0 & 0 & 0 & 0 & 0 & 0 & 0 & 0 & 0 & 0 & 0 & 0 & 1 & 0 & 0 & 0 & 0 & 0 & 0 & 0 & 0 \\
 0 & 0 & 0 & 0 & 0 & 1 & 0 & 0 & 0 & 0 & 0 & 0 & 0 & \frac{1-\mu ^2}{2}  & 0 & 0 & 0 & 0 & 0 & 0 & 0 & 0 & 0 & 0 & 1 & 0 & 0 & 0 & 0 & 0 & 0 & 0 & 0 & 0 \\
 0 & 0 & 0 & 0 & 0 & 0 & 0 & 0 & 0 & 0 & 0 & 0 & 0 & 1 & 0 & 0 & 0 & 0 & 0 & 0 & 0 & 0 & 0 & 0 & 0 & 0 & 0 & 0 & 0 & 0 & 0 & 0 & 0 & 0 \\
 0 & 0 & 0 & 0 & 0 & 0 & 0 & 0 & 0 & 0 & 0 & 0 & -\frac{2}{\mu ^2-1} & 0 & 0 & 0 & 0 & 0 & 0 & 0 & 0 & 0 & 0 & 0 & 0 & 0 & 0 & 0 & 0 & 0 & 0 & 0 & 0 & 0 \\
 0 & 0 & 0 & 0 & 1 & 0 & 0 & 0 & 0 & 0 & 0 & 0 & 1 & 0 & 0 & 0 & 0 & 0 & 0 & 0 & 0 & 0 & 0 & 1 & 0 & 0 & 0 & 0 & 0 & 0 & 0 & 0 & 0 & 0 \\
 0 & 0 & 0 & 1 & 0 & 0 & 0 & 0 & 0 & 0 & 0 & 0 & 0 & 0 & 0 & 0 & 0 & 0 & 0 & 0 & 0 & 0 & 1 & 0 & 0 & 0 & 0 & 0 & 0 & 0 & 0 & 0 & 0 & 0 \\
 0 & 0 & 1 & 0 & 0 & 0 & 0 & 0 & 0 & 0 & 0 & 0 & 0 & 0 & 0 & 0 & 0 & 0 & 0 & 0 & 0 & 1 & 0 & 0 & 0 & 0 & 0 & 0 & 0 & 0 & 0 & 0 & 0 & 0 \\
 0 & 0 & 0 & 0 & 0 & 0 & 0 & 0 & 0 & 0 & 0 & 0 & 0 & 0 & 0 & 0 & 0 & 0 & 0 & 0 & 0 & 0 & 0 & 0 & 0 & 0 & 0 & 0 & 0 & 0 & 0 & 0 & -2 \mu  & 2 \mu  \\
 0 & 1 & 0 & 0 & 0 & 0 & 0 & 0 & 0 & 0 & 0 & 0 & 0 & 0 & 0 & 0 & 0 & 0 & 0 & 0 & 1 & 0 & 0 & 0 & 0 & 0 & 0 & 0 & 0 & 0 & 0 & 0 & 0 & 0 \\
 -1 & 0 & 0 & 0 & 0 & 0 & 0 & 0 & 0 & 0 & 0 & 1 & 0 & 0 & 0 & 0 & 0 & 0 & 0 & -1 & 0 & 0 & 0 & 0 & 0 & 0 & 0 & 0 & 0 & 0 & 0 & 0 & 1 & 1 \\
 0 & 0 & 0 & 0 & 0 & 0 & 0 & 0 & 0 & 0 & 0 & 0 & 0 & 0 & 0 & 0 & 0 & 0 & 0 & 0 & 0 & 0 & 0 & 0 & 0 & 0 & 0 & 0 & 0 & 0 & 0 & 0 & 2 & 2 \\
 1 & 0 & 0 & 0 & 0 & 0 & 0 & 0 & 0 & 0 & 0 & 1 & 0 & 0 & 0 & 0 & 0 & 0 & 0 & 1 & 0 & 0 & 0 & 0 & 0 & 0 & 0 & 0 & 0 & 0 & 0 & 0 & 1 & 1 \\
 0 & 0 & 0 & 0 & 0 & 0 & 0 & 0 & 0 & 0 & 1 & 0 & 0 & 0 & 0 & 0 & 0 & 0 & 0 & 0 & 0 & 0 & 0 & 0 & 0 & 0 & 0 & 0 & 0 & 0 & 0 & 0 & 0 & 0 \\
 0 & 0 & 0 & 0 & 0 & 0 & 0 & 0 & 0 & 1 & 0 & 0 & 0 & 0 & 0 & 0 & 0 & 0 & 0 & 0 & 0 & 0 & 0 & 0 & 0 & 0 & 0 & 0 & 0 & 0 & 0 & 0 & 0 & 0 \\
 0 & 0 & 0 & 0 & 0 & 0 & 0 & 0 & 0 & 0 & 0 & 0 & 0 & 0 & 0 & 0 & 0 & 0 & 0 & 0 & 0 & 0 & 0 & 0 & 0 & 0 & 0 & 0 & 1 & 0 & 1 & 0 & 0 & 0 \\
\end{array}
\right)}
\end{equation*}
\endgroup

While the matrix $\mat{\tilde{P}}_x$ constructed in section \ref{sec:zeromodes} takes the form
\begingroup
\setlength{\arraycolsep}{2pt}
\begin{equation*}
\mat{\tilde{P}}_x = \small{
  \left(
    \begin{array}{cccccccccccccccccccccccccccccccccc}
     0 & 0 & 0 & 0 & 0 & 0 & 0 & 0 & 0 & 0 & 0 & 0 & 0 & 0 & 0 & 0 & 0 & 0 & 0 & 0 & 0 & 0 & 0 & 0 & 0 & 0 & 0 & 0 & 0 & 0 & 0 & 0 & 0 & 0 \\
     0 & 0 & 0 & 0 & 0 & 0 & 0 & 0 & 0 & 0 & -1 & 0 & 0 & 0 & 0 & 0 & 0 & 0 & 0 & 0 & 0 & 0 & 0 & 0 & 0 & 0 & 0 & 0 & 0 & 0 & 0 & 0 & 0 & 0 \\
     0 & 0 & 0 & 0 & 0 & 0 & 0 & 0 & 0 & 0 & 0 & 0 & 0 & 0 & 0 & 0 & 0 & 0 & 0 & 0 & 0 & 0 & 0 & 0 & 0 & 0 & 0 & 0 & 0 & 0 & 0 & 0 & 0 & 0 \\
     0 & 0 & 0 & 0 & 0 & 0 & 0 & 0 & 0 & 0 & 0 & 0 & 0 & 0 & 0 & 0 & 0 & 0 & 0 & 0 & 0 & 0 & 0 & 0 & 0 & 0 & 0 & 0 & 0 & 0 & 0 & 0 & 0 & 0 \\
     0 & 0 & 0 & 0 & 0 & 0 & 0 & 0 & 0 & 0 & 0 & -1 & 0 & 0 & 0 & 0 & 0 & 0 & 0 & 0 & 0 & 0 & 0 & 0 & 0 & 0 & 0 & 0 & 0 & 0 & 0 & 0 & 0 & 0 \\
     0 & 0 & 0 & 0 & 0 & 0 & 0 & 0 & 0 & 0 & 0 & 0 & 0 & 0 & 0 & 0 & 0 & 0 & 0 & 0 & 0 & 0 & 0 & 0 & 0 & 0 & 0 & 0 & 0 & 0 & 0 & 0 & 0 & 0 \\
     0 & 0 & 0 & 0 & 0 & 0 & 0 & 0 & 0 & 0 & 0 & 0 & 0 & 0 & 0 & 0 & 0 & 0 & 0 & 0 & 0 & 0 & 0 & 0 & 0 & 0 & 0 & 0 & 0 & 0 & 0 & 0 & 0 & 0 \\
     0 & 0 & 0 & 0 & 0 & -1 & 0 & 0 & 0 & 0 & 0 & 0 & 0 & 0 & 0 & 0 & 0 & 0 & 0 & 0 & 0 & 0 & 0 & 0 & 0 & 0 & 0 & 0 & 0 & 0 & 0 & 0 & 0 & 0 \\
     0 & 0 & 0 & 0 & 0 & 0 & 0 & -1 & 0 & 0 & 0 & 0 & 0 & 0 & 0 & 0 & 0 & 0 & 0 & 0 & 0 & 0 & 0 & 0 & 0 & -1 & 0 & 0 & 0 & 0 & 0 & 0 & 0 & 0 \\
     0 & 0 & 0 & 0 & 0 & 0 & 0 & 0 & 0 & 0 & 0 & 0 & 0 & 0 & 0 & 0 & 0 & 0 & 0 & 0 & 0 & 0 & 0 & 0 & 0 & 0 & 0 & 0 & 0 & 0 & 0 & 0 & 0 & 0 \\
     0 & 0 & 0 & 0 & 0 & 0 & 0 & 0 & 0 & 0 & 0 & 0 & 0 & 0 & 0 & 0 & 0 & 0 & 0 & 0 & 0 & 0 & 0 & 0 & 0 & 0 & 0 & 0 & 0 & 0 & 0 & 0 & 0 & 0 \\
     0 & 0 & 0 & 0 & 0 & 0 & 0 & 0 & 0 & 1 & 0 & 0 & 0 & 0 & 0 & 0 & 0 & 0 & 0 & 0 & 0 & 0 & 0 & 0 & 0 & 0 & 0 & 0 & 0 & 0 & 0 & 0 & 0 & 0 \\
     0 & 0 & 0 & 0 & 0 & 0 & 1 & 0 & 0 & 0 & 0 & 0 & 0 & 0 & 0 & 0 & 0 & 0 & 0 & 0 & 0 & 0 & 0 & 0 & 0 & 0 & 0 & 0 & 0 & 0 & 0 & 0 & 0 & 0 \\
     0 & 0 & -1 & 0 & 0 & 0 & 0 & 0 & 0 & 0 & 0 & 0 & 0 & 0 & 0 & 0 & 0 & 0 & 0 & 0 & 0 & -\mu ^2 & 0 & 0 & 0 & 0 & 0 & 0 & 0 & 0 & 0 & 0 & 0 & 0 \\
     0 & 0 & 0 & 0 & 0 & 0 & 0 & 0 & 0 & 0 & 0 & 0 & 0 & 0 & 0 & 0 & 0 & 0 & 0 & 0 & 0 & 0 & 0 & 0 & 0 & 0 & 0 & 0 & 0 & -\mu ^2 & 0 & 0 & 0 & 0 \\
     0 & 0 & 0 & 1 & 0 & 0 & 0 & 0 & 0 & 0 & 0 & 0 & 0 & 0 & 0 & 0 & 0 & 0 & 0 & 0 & 0 & 0 & -\mu ^2 & 0 & 0 & 0 & 0 & 0 & 0 & 0 & 0 & 0 & 0 & 0 \\
     0 & 0 & 0 & 0 & 0 & 0 & 0 & 0 & -1 & 0 & 0 & 0 & 0 & 0 & 0 & 0 & 0 & 0 & 0 & 0 & 0 & 0 & 0 & 0 & 0 & 0 & -1 & 0 & 0 & 0 & 0 & 0 & 0 & 0 \\
     0 & 0 & 0 & 0 & 0 & 0 & 0 & 0 & 0 & 0 & 0 & 0 & 0 & 0 & 0 & 0 & 0 & 0 & 0 & 0 & 0 & 0 & 0 & 0 & 0 & 0 & -1 & 0 & 0 & 0 & 0 & 0 & 0 & 0 \\
     0 & 0 & 0 & 0 & 0 & 0 & 0 & 0 & 1 & 0 & 0 & 0 & 0 & 0 & 0 & 0 & 0 & 0 & 0 & 0 & 0 & 0 & 0 & 0 & 0 & 0 & 0 & 0 & 0 & 0 & 0 & 0 & 0 & 0 \\
     0 & 0 & 0 & 0 & 0 & 0 & 0 & 0 & 0 & 0 & 0 & 0 & 0 & 0 & 0 & 0 & 0 & 0 & 0 & 0 & 0 & 0 & 0 & 0 & 0 & 0 & 0 & 0 & 0 & 0 & 0 & 0 & 0 & 0 \\
     0 & 0 & 0 & 0 & 0 & 0 & 0 & 0 & 0 & 0 & 0 & 0 & 0 & 0 & 0 & 0 & 0 & 0 & 0 & 0 & 0 & 0 & 0 & 0 & 0 & 0 & 0 & 0 & 0 & 0 & 0 & 0 & 0 & 0 \\
     0 & 0 & 0 & 0 & 0 & 0 & 0 & 0 & 0 & 0 & 0 & 0 & -1 & 0 & 0 & 0 & 0 & 0 & 0 & 0 & 0 & 0 & 0 & 0 & 0 & 0 & 0 & 0 & 0 & 0 & 0 & 0 & 0 & 0 \\
    \end{array}
    \right)
}
  \end{equation*}
  \endgroup

We also have the matrices $\mat{M}^\pm$ from Eq. \eqref{eq:conj_mat}, given by
\begin{eqnarray*}
\mat{M}^- =
&\left(
\begin{array}{ccc}
 2 \text{$\delta $Vx}+\delta \phi +\text{$\delta $e}_{\text{xx}} & 0 & 0 \\
 \frac{1}{4} \left(\text{$\delta $Vy}+\text{$\delta $e}_{\text{xy}}\right) & \frac{1}{4} \left(7 \text{$\delta $Vx}+3 \delta \phi +4 \text{$\delta
   $e}_{\text{xx}}-\text{$\delta $e}_{\text{zz}}\right) & \frac{\text{$\delta $e}_{\text{yz}}}{4} \\
 \frac{1}{4} \left(-\text{$\delta $Vz}-\text{$\delta $e}_{\text{zx}}\right) & \frac{\text{$\delta $e}_{\text{yz}}}{4} & \frac{1}{4} \left(7 \text{$\delta $Vx}+3 \delta
   \phi +4 \text{$\delta $e}_{\text{xx}}-\text{$\delta $e}_{\text{yy}}\right) \\
 0 & 0 & 0 \\
 \frac{1}{4} \left(-\text{$\delta $Vy}-\text{$\delta $e}_{\text{xy}}\right) & \frac{1}{4} \left(\text{$\delta $Vx}+\delta \phi +\text{$\delta $e}_{\text{zz}}\right) &
   -\frac{\text{$\delta $e}_{\text{yz}}}{4} \\
 \frac{1}{4} \left(\text{$\delta $Vz}+\text{$\delta $e}_{\text{zx}}\right) & -\frac{\text{$\delta $e}_{\text{yz}}}{4} & \frac{1}{4} \left(\text{$\delta $Vx}+\delta \phi
   +\text{$\delta $e}_{\text{yy}}\right) \\
 0 & 0 & 0 \\
 -\frac{2 P_{\text{yz}}+Q_{\text{xyz}}+Q_{\text{yzt}}-Q_{\text{zxy}}}{3 m^2} & -\frac{P_{\text{zx}}-Q_{\text{yzy}}+2 Q_{\text{zxt}}+Q_{\text{zxx}}}{3 m^2} &
   \frac{P_{\text{xy}}-Q_{\text{xyt}}-Q_{\text{xyx}}+Q_{\text{yzz}}}{3 m^2} \\
 0 & 0 & 0 \\
\end{array}
\right. \cdots \\
\cdots&
  \begin{array}{ccc}
   0 & 0 & 0 \\
   \frac{1}{4} \left(2 \text{$\delta $Vz}+\text{$\delta $e}_{\text{zx}}\right) & \frac{1}{4} \left(\text{$\delta $Vx}+\text{$\delta $e}_{\text{xx}}-\text{$\delta
     $e}_{\text{zz}}\right) & \frac{\text{$\delta $e}_{\text{yz}}}{4} \\
   -\frac{\text{$\delta $Vy}}{4} & \frac{\text{$\delta $e}_{\text{yz}}}{4} & \frac{1}{4} \left(\text{$\delta $Vx}+\text{$\delta $e}_{\text{xx}}-\text{$\delta
     $e}_{\text{yy}}\right) \\
   2 \text{$\delta $Vx}+\delta \phi +\text{$\delta $e}_{\text{xx}} & 0 & 0 \\
   \frac{1}{4} \left(-2 \text{$\delta $Vz}-\text{$\delta $e}_{\text{zx}}\right) & \frac{1}{4} \left(7 \text{$\delta $Vx}+4 \delta \phi +3 \text{$\delta
     $e}_{\text{xx}}+\text{$\delta $e}_{\text{zz}}\right) & -\frac{\text{$\delta $e}_{\text{yz}}}{4} \\
   \frac{\text{$\delta $Vy}}{4} & -\frac{\text{$\delta $e}_{\text{yz}}}{4} & \frac{1}{4} \left(7 \text{$\delta $Vx}+4 \delta \phi +3 \text{$\delta
     $e}_{\text{xx}}+\text{$\delta $e}_{\text{yy}}\right) \\
   0 & 0 & 0 \\
   \frac{P_{\text{x}}+P_{\text{xx}}+P_{\text{yy}}+Q_{\text{zxz}}}{3 m^2} & -\frac{P_{\text{z}}+2 P_{\text{zx}}-Q_{\text{yzy}}+Q_{\text{zxt}}}{3 m^2} & \frac{2
     P_{\text{xy}}+P_{\text{y}}+Q_{\text{xyt}}+Q_{\text{yzz}}}{3 m^2} \\
   0 & 0 & 0 \\
  \end{array}
  \cdots \\
\cdots& \left.
  \begin{array}{ccc}
   0 & 0 & 0 \\
   -\frac{\text{$\delta $Vz}}{4} & \frac{1}{4} \left(P_{\text{z}}+P_{\text{zx}}-Q_{\text{zxt}}-Q_{\text{zxx}}\right) & \frac{1}{4} \left(\text{$\delta $Vz}-\text{$\delta
     $e}_{\text{zx}}\right) \\
   \frac{1}{4} \left(2 \text{$\delta $Vy}+\text{$\delta $e}_{\text{xy}}\right) & \frac{1}{4} \left(-P_{\text{xy}}-P_{\text{y}}-2 Q_{\text{xyt}}-Q_{\text{xyx}}\right) &
     \frac{1}{4} \left(\text{$\delta $e}_{\text{xy}}-\text{$\delta $Vy}\right) \\
   0 & 0 & 0 \\
   \frac{\text{$\delta $Vz}}{4} & \frac{1}{4} \left(-P_{\text{z}}-P_{\text{zx}}+Q_{\text{zxt}}+Q_{\text{zxx}}\right) & \frac{1}{4} \left(\text{$\delta
     $e}_{\text{zx}}-\text{$\delta $Vz}\right) \\
   \frac{1}{4} \left(-2 \text{$\delta $Vy}-\text{$\delta $e}_{\text{xy}}\right) & \frac{1}{4} \left(P_{\text{xy}}+P_{\text{y}}+2 Q_{\text{xyt}}+Q_{\text{xyx}}\right) &
     \frac{1}{4} \left(\text{$\delta $Vy}-\text{$\delta $e}_{\text{xy}}\right) \\
   2 \text{$\delta $Vx}+\delta \phi +\text{$\delta $e}_{\text{xx}} & 0 & 0 \\
   -\frac{P_{\text{x}}+P_{\text{xx}}+P_{\text{zz}}-Q_{\text{xyy}}}{3 m^2} & 2 \text{$\delta $Vx}+\delta \phi +\text{$\delta $e}_{\text{xx}} &
     \frac{P_{\text{yy}}+P_{\text{zz}}-Q_{\text{xyy}}+Q_{\text{zxz}}}{3 m^2} \\
   0 & 0 & 2 \text{$\delta $Vx}+\delta \phi +\text{$\delta $e}_{\text{xx}} \\
  \end{array}
  \right)
\end{eqnarray*}

\begin{eqnarray*}
  \mat{M}^+ =
  &\left(
    \begin{array}{ccc}
     2 \text{$\delta $Vx}-\delta \phi -\text{$\delta $e}_{\text{xx}} & 0 & 0 \\
     \frac{1}{4} \left(\text{$\delta $e}_{\text{xy}}-\text{$\delta $Vy}\right) & \frac{1}{4} \left(7 \text{$\delta $Vx}-3 \delta \phi -4 \text{$\delta
       $e}_{\text{xx}}+\text{$\delta $e}_{\text{zz}}\right) & -\frac{\text{$\delta $e}_{\text{yz}}}{4} \\
     \frac{1}{4} \left(\text{$\delta $Vz}-\text{$\delta $e}_{\text{zx}}\right) & -\frac{\text{$\delta $e}_{\text{yz}}}{4} & \frac{1}{4} \left(7 \text{$\delta $Vx}-3 \delta
       \phi -4 \text{$\delta $e}_{\text{xx}}+\text{$\delta $e}_{\text{yy}}\right) \\
     0 & 0 & 0 \\
     \frac{1}{4} \left(\text{$\delta $e}_{\text{xy}}-\text{$\delta $Vy}\right) & \frac{1}{4} \left(-\text{$\delta $Vx}+\delta \phi +\text{$\delta $e}_{\text{zz}}\right) &
       -\frac{\text{$\delta $e}_{\text{yz}}}{4} \\
     \frac{1}{4} \left(\text{$\delta $Vz}-\text{$\delta $e}_{\text{zx}}\right) & -\frac{\text{$\delta $e}_{\text{yz}}}{4} & \frac{1}{4} \left(-\text{$\delta $Vx}+\delta \phi
       +\text{$\delta $e}_{\text{yy}}\right) \\
     0 & 0 & 0 \\
     -\frac{2 P_{\text{yz}}-Q_{\text{xyz}}+Q_{\text{yzt}}+Q_{\text{zxy}}}{3 m^2} & \frac{P_{\text{zx}}+Q_{\text{yzy}}+2 Q_{\text{zxt}}-Q_{\text{zxx}}}{3 m^2} &
       \frac{-P_{\text{xy}}+Q_{\text{xyt}}-Q_{\text{xyx}}+Q_{\text{yzz}}}{3 m^2} \\
     0 & 0 & 0 \\
    \end{array}
  \right. \cdots \\
  \cdots& 
  \begin{array}{ccc}
    0 & 0 & 0 \\
    \frac{1}{4} \left(\text{$\delta $e}_{\text{zx}}-2 \text{$\delta $Vz}\right) & \frac{1}{4} \left(-\text{$\delta $Vx}+\text{$\delta $e}_{\text{xx}}-\text{$\delta
      $e}_{\text{zz}}\right) & \frac{\text{$\delta $e}_{\text{yz}}}{4} \\
    \frac{\text{$\delta $Vy}}{4} & \frac{\text{$\delta $e}_{\text{yz}}}{4} & \frac{1}{4} \left(-\text{$\delta $Vx}+\text{$\delta $e}_{\text{xx}}-\text{$\delta
      $e}_{\text{yy}}\right) \\
    2 \text{$\delta $Vx}-\delta \phi -\text{$\delta $e}_{\text{xx}} & 0 & 0 \\
    \frac{1}{4} \left(\text{$\delta $e}_{\text{zx}}-2 \text{$\delta $Vz}\right) & \frac{1}{4} \left(7 \text{$\delta $Vx}-4 \delta \phi -3 \text{$\delta
      $e}_{\text{xx}}-\text{$\delta $e}_{\text{zz}}\right) & \frac{\text{$\delta $e}_{\text{yz}}}{4} \\
    \frac{\text{$\delta $Vy}}{4} & \frac{\text{$\delta $e}_{\text{yz}}}{4} & \frac{1}{4} \left(7 \text{$\delta $Vx}-4 \delta \phi -3 \text{$\delta
      $e}_{\text{xx}}-\text{$\delta $e}_{\text{yy}}\right) \\
    0 & 0 & 0 \\
    \frac{-P_{\text{x}}+P_{\text{xx}}+P_{\text{yy}}-Q_{\text{zxz}}}{3 m^2} & -\frac{-P_{\text{z}}+2 P_{\text{zx}}+Q_{\text{yzy}}+Q_{\text{zxt}}}{3 m^2} & -\frac{-2
      P_{\text{xy}}+P_{\text{y}}-Q_{\text{xyt}}+Q_{\text{yzz}}}{3 m^2} \\
    0 & 0 & 0 \\
   \end{array}
  \cdots \\
  \cdots& \left.
    \begin{array}{ccc}
     0 & 0 & 0 \\
     \frac{\text{$\delta $Vz}}{4} & \frac{1}{4} \left(P_{\text{z}}-P_{\text{zx}}+Q_{\text{zxt}}-Q_{\text{zxx}}\right) & \frac{1}{4} \left(\text{$\delta $Vz}+\text{$\delta
       $e}_{\text{zx}}\right) \\
     \frac{1}{4} \left(\text{$\delta $e}_{\text{xy}}-2 \text{$\delta $Vy}\right) & \frac{1}{4} \left(P_{\text{xy}}-P_{\text{y}}+2 Q_{\text{xyt}}-Q_{\text{xyx}}\right) &
       \frac{1}{4} \left(-\text{$\delta $Vy}-\text{$\delta $e}_{\text{xy}}\right) \\
     0 & 0 & 0 \\
     \frac{\text{$\delta $Vz}}{4} & \frac{1}{4} \left(P_{\text{z}}-P_{\text{zx}}+Q_{\text{zxt}}-Q_{\text{zxx}}\right) & \frac{1}{4} \left(\text{$\delta $Vz}+\text{$\delta
       $e}_{\text{zx}}\right) \\
     \frac{1}{4} \left(\text{$\delta $e}_{\text{xy}}-2 \text{$\delta $Vy}\right) & \frac{1}{4} \left(P_{\text{xy}}-P_{\text{y}}+2 Q_{\text{xyt}}-Q_{\text{xyx}}\right) &
       \frac{1}{4} \left(-\text{$\delta $Vy}-\text{$\delta $e}_{\text{xy}}\right) \\
     2 \text{$\delta $Vx}-\delta \phi -\text{$\delta $e}_{\text{xx}} & 0 & 0 \\
     -\frac{-P_{\text{x}}+P_{\text{xx}}+P_{\text{zz}}+Q_{\text{xyy}}}{3 m^2} & 2 \text{$\delta $Vx}-\delta \phi -\text{$\delta $e}_{\text{xx}} &
       -\frac{P_{\text{yy}}+P_{\text{zz}}+Q_{\text{xyy}}-Q_{\text{zxz}}}{3 m^2} \\
     0 & 0 & 2 \text{$\delta $Vx}-\delta \phi -\text{$\delta $e}_{\text{xx}} \\ 
    \end{array}
    \right)
\end{eqnarray*}

\clearpage

\addcontentsline{toc}{section}{Bibliography}
\bibliography{refs}

\end{document}